\documentclass[]{aa} 

\usepackage{graphicx}
\usepackage{natbib}
\usepackage{txfonts}
\usepackage{academicons}
\usepackage{hyperref}
\usepackage{xcolor}
\usepackage{ulem}

\usepackage{soul}

\begin{document} 
   \title{ The GRAVITY Young Stellar Object Survey}
   \subtitle{VI. Mapping the variable inner disk of HD\,163296 at sub-au scales}

   \author{GRAVITY Collaboration (\thanks{GRAVITY is developed in a
collaboration by the Max Planck Institute for Extraterrestrial Physics,
LESIA of Paris Observatory and IPAG of Université Grenoble Alpes / CNRS,
the Max Planck Institute for Astronomy, the University of Cologne, the
Centro de Astrofísica e Gravitação and the European
Southern Observatory.}): {J. Sanchez-Bermudez}
          \inst{1,2}  \and A. Caratti o Garatti \inst{2,3,4} \and R. Garcia Lopez \inst{2,3,4} \and K. Perraut\inst{5} \and L. Labadie\inst{6} \and M. Benisty\inst{5, 9} \and W. Brandner\inst{2} \and  C. Dougados\inst{5} \and P.J.V. Garcia\inst{7,8} \and Th. Henning\inst{2} \and L. Klarmann\inst{2} \and A. Amorim\inst{7} \and M. Baub\"ock\inst{11} \and J.P. Berger\inst{5} \and J.B. Le Bouquin\inst{5} \and P. Caselli\inst{11} \and Y. Cl\'enet\inst{10} \and V. Coud\'e du Foresto\inst{10} \and P.T. de Zeeuw\inst{11} \and A. Drescher\inst{11} \and G. Duvert\inst{5} \and A. Eckart\inst{6} \and F. Eisenhauer\inst{11}  \and M. Filho\inst{7,8} \and F. Gao\inst{11} \and E. Gendron\inst{10} \and R. Genzel\inst{11} \and S. Gillessen\inst{11} \and R. Grellmann\inst{6} \and G. Heissel\inst{10} \and M. Horrobin\inst{6} \and Z. Hubert\inst{5} \and A. Jim\'enez-Rosales\inst{11} \and L. Jocou\inst{5} \and P. Kervella\inst{10} \and S. Lacour\inst{10} \and V. Lapeyr\`ere\inst{10} \and P. L\'ena\inst{10} \and T. Ott\inst{11} \and T. Paumard\inst{10} \and G. Perrin\inst{10} \and  J. E. Pineda\inst{11} \and G. Rodr\'iguez-Coira\inst{10} \and G. Rousset\inst{10} \and D. M. Segura-Cox\inst{11} \and J. Shangguan\inst{11} \and T. Shimizu\inst{11} \and J. Stadler\inst{11} \and O. Straub\inst{11} \and C. Straubmeier\inst{6} \and E. Sturm\inst{11} \and E. van Dishoeck\inst{11} \and F. Vincent\inst{10}  \and S.D. von Fellenberg\inst{11} \and F. Widmann\inst{11}   \and J. Woillez\inst{12}}

   \institute{Instituto de Astronom\'ia, Universidad Nacional Aut\'onoma de M\'exico, Apdo. Postal 70264, Ciudad de M\'exico, 04510, M\'exico\\
              \email{joelsb@astro.unam.mx}
\and
             Max-Planck-Institut f\"ur Astronomie, K\"{o}nigstuhl 17, D-69117 Heidelberg, Germany
             \and
             Dublin Institute for Advanced Studies, 31 Fitzwilliam Place, D02\,XF86 Dublin, Ireland
             \and
              School of Physics, University College Dublin, Belfield, Dublin 4, Ireland
             \and
             Univ. Grenoble Alpes, CNRS, IPAG, F-38000 Grenoble, France
             \and 
              I. Physikalisches Institut, Universität zu Köln, Zülpicher Str. 77, 50937, Köln, Germany
             \and
Universidade do Porto - Faculdade de Engenharia, Rua Dr. Roberto Frias, 4200-465 Porto, Portugal
\and
CENTRA, Instituto Superior Tecnico, Av. Rovisco Pais, 1049-001 Lisboa, Portugal
\and
Unidad Mixta Internacional Franco-Chilena de Astronomía (CNRS UMI 3386), Departamento de Astronomía, Universidad de Chile, Camino El Observatorio 1515, Las Condes, Santiago, Chile
\and
LESIA, Observatoire de Paris, PSL Research University, CNRS, Sorbonne Universit\'es, UPMC Univ. Paris 06, Univ. Paris Diderot, Sorbonne Paris Cit\'e, France
\and
Max Planck Institute for Extraterrestrial Physics, Giessenbachstrasse, 85741 Garching bei M\"{u}nchen, Germany
\and European Southern Observatory, Karl-Schwarzschild-Str. 2, 85748
Garching, Germany
             }

  \date{Received, accepted}

 
  \abstract
   {Protoplanetary disks drive some of the formation process (e.g., accretion, gas dissipation, formation of structures, etc.) of stars and planets. Understanding such physical processes is one of the main astrophysical questions. HD\,163296 is an interesting young stellar object for which infrared and sub-millimeter observations have shown a prominent circumstellar disk with gaps plausibly created by forming planets. }
   {This study aims at characterizing the morphology of the inner disk in \object{HD\,163296} with 
   multi-epoch near-infrared interferometric observations performed with GRAVITY at the Very Large Telescope Interferometer (VLTI). Our goal is to depict the $K-$band ($\lambda_0 \sim$ 2.2~$\mu$m) structure of the inner rim with milliarcsecond (sub-au) angular resolution. Our data is complemented with archival PIONIER ($H-$band; $\lambda_0 \sim$ 1.65~$\mu$m) data of the source.}
   {We performed a Gradient Descent parametric model fitting to recover the sub-au morphology of our source.}
   {Our analysis shows the existence of an asymmetry in the disk surrounding the central star of HD\,163296. We confirm variability of the disk structure in the inner $\sim$2 mas (0.2 au). While variability of the inner disk structure in this source has been suggested by previous interferometric studies, this is the first time that it is confirmed in the $H$- and $K$-bands by using a complete analysis of the closure phases and squared visibilities over several epochs. Because of the separation from the star, position changes, and persistence of this asymmetric structure on timescales of several years, we argue that it is a dusty feature (e.g., a vortex or dust clouds), probably, made by a mixing of sillicate and carbon dust and/or refractory grains, in-homogeneously distributed above the mid-plane of the disk.}
   {}

   \keywords{young stellar objects --
                accretion disks --
                infrared interferometry
               }

   \maketitle
%

   \section{Introduction}

The formation and evolution of protoplanetary disks is crucial in the formation process of stars and planets. They are key laboratories for magneto-hydrodynamic, radiative and astro-chemical processes. Understanding the processes at play in these dust- and gas-rich circumstellar disks is one of the main science cases of several current observing facilities. ALMA \citep{ALMA2015} and adaptive-optics assisted imagers, like SPHERE \citep{Beuzit2019}, have revealed the complexity and the diversity of the outer parts of these disks (e.\,g., $\sim$~20-500\,au). Most of them exhibit diverse features as gaps, rings, spirals, vortex, and shadows \citep{Andrews2018,Long2018,Benisty2015,deBoer2016,Pohl2017,Benisty2017,Avenhaus2018} whose origin is still a matter of debate. Knowing the disk properties at different spatial scales and, in particular, in the innermost regions close to, or within the dust sublimation front, is critical to understand the conditions for planet formation and migration in proto-stellar disks around young stars \citep{Flock2019}.


Zooming into these innermost regions requires angular resolution down to a few milliarcseconds (mas) or less, which is made only possible with optical long-baseline interferometry. Several statistical studies of the dust inner rims of young stellar objects (YSOs) have been conducted using instruments at the Very Large Telescope Interferometer (VLTI) in $H$ ($\lambda_0 \sim$ 1.65~$\mu$m)  and $K$ ($\lambda_0 \sim$ 2.2~$\mu$m) bands \citep{Lazareff2017, Anthonioz2015, Perraut2019}. Recent reconstructed images of 15 Herbig stars in the $H-$band \citep{Kluska2020} show that 60\% exhibits disk asymmetries at a few au scale, and 27\% has a non-centrosymmetric morphology that can be due to a non-axisymmetric and/or variable environment \citep[][]{Chen2019}.

 \begin{figure*}[tp]
\textbf{HD\,163296 - Best-fit azimuthally modulated ring model}\par\medskip
  \centering
  \includegraphics[width=18 cm]{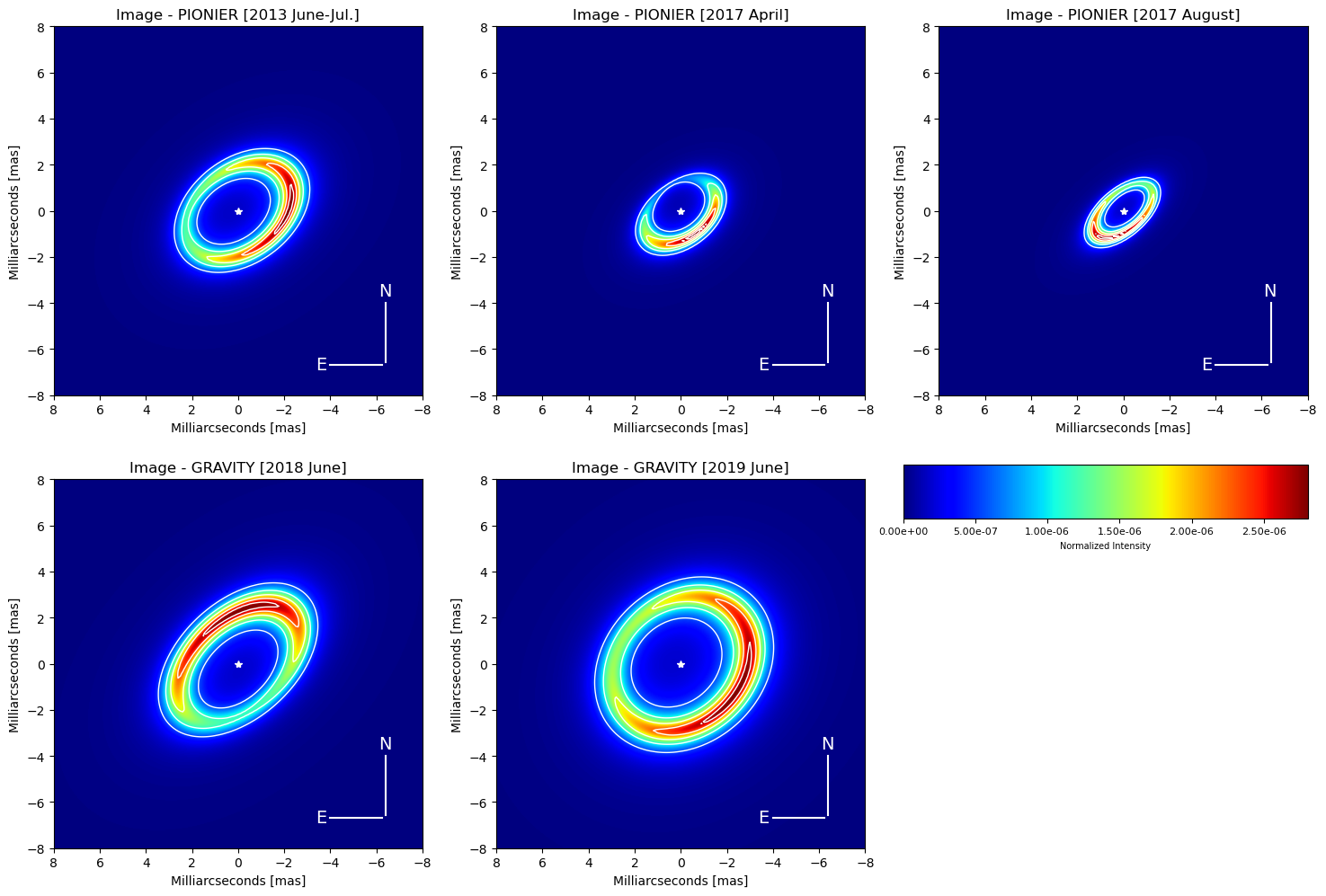}
\caption{Panels show a mean image of the best-fit azimuthally modulated ring model described in Sec. \ref{sec:ring_model}. Instrument and epochs are displayed on each panel. These  mean images were created from the best-fit models of  the individual wavelengths per epoch. All panels are plotted with the same colormap and normalized flux scale. The white contours in the images represent the 20\%, 40\%, 60\%, 80\% and 95\% of the peak's emission. The white star in the center of the images traces the position of the star in the model.}
\label{fig:mean_gravity_images}
\end{figure*}

The Herbig Ae star \object{HD\,163296} has a mass of 2.23 $\pm$ 0.22 M$_{\odot}$ \citep{Alecian_2013} and is located at a distance of 101 $\pm$ 1.2 pc \citep{Vioque_2018}. This makes it one of the closest Herbig stars to us. Its protoplanetary disk has been well studied in a wide spectral range, from the optical to the millimetric part. The ALMA DSHARP project \citep{Andrews2018} reveals several rings around this star and the authors report detailed azimuthal asymmetries in the outer disk \citep{Huang2018,Isella2018}. Such features may be produced by planets, whose presence is also suggested by deviations from Keplerian rotational motions \citep{Teague2018, Teague_2019,Pinte2018}. Multi-epoch study of HD~163296 in the optical with coronagraphic imagers revealed a temporally variable, non-azimuthally symmetric illumination of the outer disk \citep{Rich_2019, Rich_2020}, while $J$-band polarimetric observations detected an off-center ring around HD~163296, which appears to be caused by scattering of the upper layers  \citep{Monnier2017}. Also, it is one of the few Herbig stars to drive a prominent jet \citep{Ellerbroek_2014}.

Optical interferometric studies of HD~163296 show the presence of an elongated dusty disk with a diameter of $\sim$5\,mas with an inclination of $\sim$45\degr\ and a position angle of $\sim$130\degr\ \citep{Lazareff2017, Setterholm_2018,Perraut2019}. In these works, the closure phases show a deviation from zero on the long baselines, which hints at an asymmetric structure or clumpy emission in the innermost disk region. In particular, \citet{Setterholm_2018} show strong closure phase variations for the longest baselines sampled with CLIMB-CHARA. These authors suggest the presence of an asymmetric emission at scales smaller than $\sim$2\,mas ($
\sim$0.2\,au).

Recently, \citet{Kluska2020} reconstructed an image of HD\,163296 using PIONIER-VLTI ($H-$band). These authors discussed how a possible asymmetry located at the inner edge of the disk would move during the observing time spanning the interferometric observations, and how this could be affecting the interpretation of the observed morphology. For example, a body in Keplerian rotation at a radius of 3\,mas (or 0.3\,au) would complete a full orbit in $\sim$40 days, assuming a central mass of 2.23\,M$_{\odot}$. This puts strong constraints in the way interferometric data could be combined for imaging. In contrast, it also motivates monitoring campaigns to depict the nature of this asymmetry. Additionally, \citet{Kobus_2020} suggested symmetric variability at sub-au scales based on the relative changes of the squared visibilities observed in PIONIER and AMBER ($\lambda_0$ = 2.2 $\mu$m) data. However, these analyses do not include the characterization of the closure phases. Therefore, the asymmetry of the source could not be fully depicted. More recently, \citet{Varga_2020} detected an asymmetric structure using MATISSE data in the L-band ($\lambda_0$ = 3.8 $\mu m$), which produces a brighter side on the extended emission in the inner 4 mas (0.4 au) of the source. This asymmetric brightness distribution has been identified before by \citet{Lazareff2017} and \citet{Kluska2020}. However, the interesting aspect is that the position angle of the asymmetry detected with MATISSE appears to be significantly different from the one reported in the PIONIER observations. Thus, this suggests a varying morphology of the inner disk. However, the limited MATISSE data on large baselines did not allow a proper characterization of the asymmetry and its supposedly variable nature.

This work is a further step in the analysis of the inner $\leq$ 1\,au ($\leq$ 10 mas) of HD\,163296. We aim at depicting the morphology of the inner disk by using multi-epoch interferometric observations obtained with GRAVITY-VLTI in $K$-band and archival PIONIER $H-$band data. Our specific goal is to characterize the predicted asymmetric structure in a systematic way across the $H$ and $K$ bands. For this purpose, we present a geometrical model to the squared visibilities (V$^2$) and closure phases (CPs). The manuscript is divided as follows: Sec.\ref{sec:Data} presents the interferometric data used for this work; Sec. \ref{sec:ring_model} describes our model and Sec.\ref{sec:Results} our results; in Sec. \ref{sec:Discussion}, we present our discussion and; in Sec.\,\ref{sec:Conclusions} our conclusions are presented.

\section{Observations and Data Reduction \label{sec:Data}}
\subsection{GRAVITY data}
HD\,163296 was observed with
GRAVITY \citep{Eisenhauer_2008, Eisenhauer_2011, Eisenhauer_2017} during 2018 and 2019 as part of the Young Stellar Object (YSO) Guaranteed Time Observations (GTO). The log of the observations is reported in Table\,\ref{tab:observations}. The observations were done
using the highest spectral resolution (R $\sim$ 4000) of the
instrument, deploying the science source as Fringe Tracker \citep[R $\sim$ 22, sampling $\sim$1 kHz;][]{Gillessen_2010}. This allows us to stabilize the fringes of the science beam combiner for up to several tens of seconds. All the data were recorded using the 1.8 m Auxiliary Telescopes (ATs) with the intermediate array (D0-G2-J3-K0)\footnote{The GRAVITY observations used in this study are based on the data collected through the ESO programs: 0101.C-0311(A) and 103.C-0347(A). The PIONIER observations used in this study are based on the data collected through the ESO programs: 99.C-0546(B) and 99.C-0546(C).}. The interferometric observables (squared visibilities and closure
phases) were obtained using the instrument's
data reduction software provided by ESO \citep{Lapeyrere_2014}.

Absolute calibration of the science data is implemented through the observations of point-like sources interleaved with the science sequences. The data reduction software estimates the instrumental transfer function by correcting the observed calibrator
visibilities with the theoretical ones according to the estimated
angular size of the calibrator. Finally, the algorithm corrects the science raw observables by the estimated transfer function. Before analyzing the data, all squared visibilities (V$^2$) with a signal-to-noise ratio (SNR) lower than five and closure phases (CPs) with uncertainties above 40$^{\circ}$ were
discarded.

Post-processing of the data was done with custom-made python routines. In this step, science data were wavelength re-binned from the native $\sim$1700 channels of the high-resolution data sets down to 8 spectral channels across the $K-$band, which increased the signal-to-noise for our analysis of the continuum. We account for calibration errors, while keeping the SNR statistics of the different data sets. We obtained average error bars of $\sigma_{V
^2}$ = 0.01 and $\sigma_{CPs}$ = 0.3$^{\circ}$ for the 2019 data and $\sigma_{V
^2}$ = 0.01 and $\sigma_{CPs}$ = 0.8$^{\circ}$ for the 2018 data, respectively. Our analysis is based on the exploitation of data from the science channel of GRAVITY and not from the fringe-tracker channel of the instrument. Detailed analysis of emission lines like Br$\gamma$ were discarded and are the subject of a forth-coming paper. GRAVITY u-v planes are included in Figure \ref{fig:uv_planes} in the Appendix. 

\subsection{Ancillary Data}

To complement the analysis and results obtained from our GRAVITY data, we used the available PIONIER-VLTI data sets of HD\,163296 taken during 2013, and 2017, which are included into the Optical Interferometry Data Base (OIDB\footnote{http://oidb.jmmc.fr/index.html}) hosted by the Jean-Marie Mariotti Center (JMMC). These data are reduced and calibrated in a consistent way, which allows us to compare and merge them directly. PIONIER data are of low-spectral resolution, sampling three and six spectral channels across the $H-$band ($\lambda_{0}$= 1.65 $\mu$m) in 2013, and 2017, respectively. The log of these observations are also listed in Table\,\ref{tab:observations} in the Appendix. 

\pagebreak
\newpage

\begin{table*}[]
\caption[]{PIONIER - Best-fit parameters of the azimuthally modulated ring model}
\label{tab:param_ring_2013}
\centering
\rotatebox{0}{
\begin{tabular}{l c c c c c c } 
\hline \hline
\multicolumn{7}{c}{\textbf{2013 - June/July}} \\
\hline
Wavelengths [microns] & 1.618 & 1.7 & 1.778 \\
\hline
PA [deg]                        & 126.5 $\pm$ 3.0     & 127.2  $\pm$ 1.7    & 126.6 $\pm$ 1.6  & & & \\
$i$ [deg]                             & 44.8 $\pm$ 1.7      & 45.9 $\pm$ 1.2      & 47.7 $\pm$ 1.3  & & &   \\
c$_1$                                     & -0.17 $\pm$ 0.1     & -0.25  $\pm$ 0.08    & -0.37 $\pm$ 0.1  & & &  \\
s$_1$                                     & -0.28 $\pm$ 0.07    & -0.28 $\pm$ 0.05    & -0.37 $\pm$ 0.06  & & &  \\
a$_r$ [mas]                              & 1.75 $\pm$ 0.05      & 1.8 $\pm$ 0.04      & 1.8 $\pm$ 0.04 & & &   \\
$F_{\mathrm{s}}$                      & 0.45 $\pm$ 0.01     & 0.43 $\pm$ 0.01     & 0.41 $\pm$ 0.01   & & &  \\
$F_{\mathrm{c}}$                      & 0.53 $\pm$ 0.01     & 0.55 $\pm$ 0.01     & 0.56 $\pm$ 0.01   & & &   \\
$F_{\mathrm{h}}$                   & 0.02 $\pm$ 0.01     & 0.02 $\pm$ 0.01     & 0.03 $\pm$ 0.01     & & &  \\
$\chi^2$ & 2.2 & 2.5 & 3.5  \\
\hline
\hline
\multicolumn{7}{c}{\textbf{2017 - April}} \\
\hline
Wavelengths [microns] & 1.518 & 1.567 & 1.617 & 1.67 & 1.72 & 1.763 \\
\hline
PA [deg]                        & 131.0 $\pm$ 1.8   & 132.0  $\pm$ 1.9  & 129.3 $\pm$ 2.6  & 129.1 $\pm$ 2.0  & 124.7 $\pm$ 3.9  &  128.5 $\pm$ 3.5 \\
$i$ [deg]                             & 51.5 $\pm$ 1.2    & 50.5 $\pm$ 0.9    & 49.9 $\pm$ 0.9   & 50.1 $\pm$ 0.8   & 42.6 $\pm$ 1.5  &  45.5 $\pm$ 1.9 \\
c$_1$                                     & -0.34 $\pm$ 0.09  & -0.46  $\pm$ 0.07  & -0.52 $\pm$ 0.1  & -0.37 $\pm$ 0.1  & -1.0 $\pm$ 0.05 &  -0.85 $\pm$ 0.1 \\
s$_1$                                     & 0.6 $\pm$ 0.2     & 0.50 $\pm$ 0.14    & 0.4 $\pm$ 0.13 &  0.2 $\pm$ 0.1 & -0.29 $\pm$ 0.1  & -0.17 $\pm$ 0.2 \\
a$_r$ [mas]                              & 1.16 $\pm$ 0.06   & 1.22 $\pm$ 0.05   & 1.22 $\pm$ 0.06  &  1.22 $\pm$ 0.07 & 1.15 $\pm$ 0.02 & 1.18 $\pm$ 0.02\\
$F_{\mathrm{s}}$                       & 0.42 $\pm$ 0.01   & 0.41 $\pm$ 0.02   & 0.38 $\pm$ 0.02  &  0.36 $\pm$ 0.03 & 0.21 $\pm$ 0.02  &  0.24 $\pm$ 0.03 \\
$F_{\mathrm{c}}$                       & 0.44 $\pm$ 0.01   & 0.46 $\pm$ 0.01   & 0.47 $\pm$ 0.02  &  0.48 $\pm$ 0.02 & 0.61 $\pm$ 0.02  &  0.58 $\pm$ 0.03 \\
$F_{\mathrm{h}}$                   & 0.13 $\pm$ 0.01    & 0.13 $\pm$ 0.01    & 0.15 $\pm$ 0.02   &  0.16 $\pm$ 0.02  &  0.18 $\pm$ 0.01  & 0.18 $\pm$ 0.01 \\
$\chi^2$ & 1.2 & 1.9 & 2.8 & 3.1 & 2.9 & 3.0 \\
\hline
  \hline
\multicolumn{7}{c}{\textbf{2017 - August}} \\
\hline
Wavelengths [microns] & 1.518 & 1.567 & 1.617 & 1.67 & 1.72 & 1.763 \\
\hline
PA [deg]                        & 131.1 $\pm$ 1.4   & 130.2  $\pm$ 1.1  & 130.7 $\pm$ 1.1  & 130.9 $\pm$ 1.1  & 131.5 $\pm$ 1.0  &  132.3 $\pm$ 0.9 \\
$i$ [deg]                             & 58.9 $\pm$ 1.0    & 58.2 $\pm$ 0.82    & 59.0 $\pm$ 0.8   & 59.1 $\pm$ 0.8   & 59.5 $\pm$ 0.7  &  50.1 $\pm$ 0.8 \\
c$_1$                                     & -0.45 $\pm$ 0.2  & -0.47  $\pm$ 0.13  & -0.17 $\pm$ 0.17  & -0.09 $\pm$ 0.2  & -0.2 $\pm$ 0.15 &  0.04 $\pm$ 0.2 \\
s$_1$                                     & 0.27 $\pm$ 0.13     & 0.29 $\pm$ 0.09    & 0.26 $\pm$ 0.1 &  0.26 $\pm$ 0.1 & 0.44 $\pm$ 0.1  & 0.31 $\pm$ 0.1 \\
a$_r$ [mas]                              &0.77 $\pm$ 0.03   & 0.81 $\pm$ 0.02   & 0.8 $\pm$ 0.02  &  0.8 $\pm$ 0.02 & 0.84 $\pm$ 0.02 & 0.86 $\pm$ 0.02\\
$F_{\mathrm{s}}$                       & 0.38 $\pm$ 0.01   & 0.37 $\pm$ 0.01   & 0.34 $\pm$ 0.01  &  0.32 $\pm$ 0.01 & 0.31 $\pm$ 0.01  &  0.29 $\pm$ 0.01 \\
$F_{\mathrm{c}}$                       & 0.41 $\pm$ 0.01   & 0.44 $\pm$ 0.01   & 0.46 $\pm$ 0.01  &  0.46 $\pm$ 0.01 & 0.47 $\pm$ 0.01  &  0.49 $\pm$ 0.01 \\
$F_{\mathrm{h}}$                   & 0.21 $\pm$ 0.02    & 0.19 $\pm$ 0.01    & 0.20 $\pm$ 0.01   &  0.22 $\pm$ 0.02  &  0.22 $\pm$ 0.01  & 0.22 $\pm$ 0.01 \\ 
$\chi^2$ & 2.6 & 2.6 & 3.8 & 5.2 & 3.4 & 2.6 \\
\hline
\end{tabular}
}
\end{table*}

\clearpage
\newpage

\begin{table*}[]
\caption[]{GRAVITY - Best-fit parameters of the azimuthally modulated ring model}
\label{tab:param_ring_2018}
\centering
\rotatebox{90}{
\begin{tabular}{l c c c c c c c c } 
\hline \hline
\multicolumn{9}{c}{\textbf{2018 - June}} \\
\hline
Wavelengths [microns] & 2.016 & 2.069 & 2.122 & 2.175 & 2.228 & 2.281 & 2.333 & 2.386 \\
\hline
PA [deg]                        & 132.5 $\pm$ 0.9     & 132.5  $\pm$ 0.8    & 132.5 $\pm$ 0.7   & 133.0 $\pm$ 0.7   & 133.9 $\pm$ 0.7   & 133.0 $\pm$ 0.7  & 133.7 $\pm$ 0.7    & 134.3 $\pm$ 0.8 \\
$i$ [deg]                             & 52.5 $\pm$ 0.4      & 52.3 $\pm$ 0.4      & 52.1 $\pm$ 0.3    & 51.9 $\pm$ 0.3    & 51.8 $\pm$ 0.3    & 51.6 $\pm$ 0.3   & 51.1 $\pm$ 0.3     & 50.4 $\pm$ 0.3 \\
c$_1$                                     & 0.45 $\pm$ 0.04     & 0.41  $\pm$ 0.03    & 0.38 $\pm$ 0.03   & 0.34 $\pm$ 0.03   & 0.37 $\pm$ 0.03   & 0.36 $\pm$ 0.03  & 0.36 $\pm$ 0.03    & 0.35 $\pm$ 0.04 \\
s$_1$                                     & -0.25 $\pm$ 0.05    & -0.23 $\pm$ 0.04    & -0.2 $\pm$ 0.03   & -0.21 $\pm$ 0.03  & -0.22 $\pm$ 0.04  & -0.17 $\pm$ 0.03 & -0.21 $\pm$ 0.04   & -0.2 $\pm$ 0.05 \\
a$_r$ [mas]                              & 1.7 $\pm$ 0.03      & 1.8 $\pm$ 0.02      & 1.9 $\pm$ 0.03    & 2.0 $\pm$ 0.03    & 2.0 $\pm$ 0.03    & 2.04 $\pm$ 0.03  & 2.09 $\pm$ 0.03    & 2.15 $\pm$ 0.04 \\
$F_{\mathrm{s}}$                      & 0.33 $\pm$ 0.008     & 0.34 $\pm$ 0.006     & 0.35 $\pm$ 0.006    & 0.35 $\pm$ 0.006    & 0.35 $\pm$ 0.006 & 0.34 $\pm$ 0.006  & 0.34 $\pm$ 0.006    & 0.34 $\pm$ 0.008 \\
$F_{\mathrm{c}}$                      & 0.50 $\pm$ 0.006     & 0.52 $\pm$ 0.006     & 0.53 $\pm$ 0.005    & 0.53 $\pm$ 0.005    & 0.54 $\pm$ 0.005 & 0.54 $\pm$ 0.005  & 0.54 $\pm$ 0.005    & 0.55 $\pm$ 0.006 \\
$F_{\mathrm{h}}$                   & 0.17 $\pm$ 0.01     & 0.14 $\pm$ 0.009     & 0.12 $\pm$ 0.009    & 0.12 $\pm$ 0.008    & 0.11 $\pm$ 0.009 & 0.12 $\pm$ 0.008  & 0.12 $\pm$ 0.008    & 0.11 $\pm$ 0.009 \\
$\chi$2 & 3.0 & 8.5 & 12.2 & 6.0 & 4.5 & 5.1 & 5.5 & 3.7 \\
\hline \hline
\multicolumn{9}{c}{\textbf{2019 - June}} \\
\hline
Wavelengths [microns] & 2.016 & 2.069 & 2.122 & 2.175 & 2.228 & 2.281 & 2.333 & 2.386 \\
\hline
PA [deg]                        & 137.7 $\pm$ 2.4    & 134.4  $\pm$ 1.1  & 132.8 $\pm$ 1.7   & 132.1 $\pm$ 1.6   & 131.9 $\pm$ 1.8   & 129.1 $\pm$ 1.4   & 130.1 $\pm$ 1.4  & 128.2 $\pm$ 1.9 \\
$i$ [deg]                             & 33.8 $\pm$ 1.2     & 35.4 $\pm$ 0.5    & 34.2 $\pm$ 0.7    & 34.6 $\pm$ 0.7    & 34.4 $\pm$ 0.8    & 35.5 $\pm$ 0.7    & 35.5 $\pm$ 0.8   & 35.5 $\pm$ 1.0 \\
c$_1$                                     & -0.24 $\pm$ 0.01   & -0.28  $\pm$ 0.006 & -0.27 $\pm$ 0.01  & -0.27 $\pm$ 0.008  & -0.28 $\pm$ 0.01  & -0.29 $\pm$ 0.01  & -0.31 $\pm$ 0.01 & -0.32 $\pm$ 0.02 \\
s$_1$                                     & -0.11 $\pm$ 0.02   & -0.14 $\pm$ 0.02  & -0.14 $\pm$ 0.01  & -0.16 $\pm$ 0.02  & -0.15 $\pm$ 0.02  & -0.14 $\pm$ 0.01  & -0.15 $\pm$ 0.02 & -0.16 $\pm$ 0.02 \\
a$_r$ [mas]                              & 2.61 $\pm$ 0.05    & 2.60 $\pm$ 0.01   & 2.61 $\pm$ 0.03   & 2.69 $\pm$ 0.02   & 2.73 $\pm$ 0.02   & 2.74 $\pm$ 0.02   & 2.77 $\pm$ 0.03  & 2.80 $\pm$ 0.03 \\
$F_{\mathrm{s}}$                      & 0.44 $\pm$ 0.003    & 0.43 $\pm$ 0.002   & 0.43 $\pm$ 0.002   & 0.43 $\pm$ 0.003   & 0.42 $\pm$ 0.002    & 0.42 $\pm$ 0.002  & 0.41 $\pm$ 0.003    & 0.41 $\pm$ 0.004 \\
$F_{\mathrm{c}}$                     & 0.56 $\pm$ 0.003    & 0.57 $\pm$ 0.002   & 0.57 $\pm$ 0.002   & 0.57 $\pm$ 0.002   & 0.58 $\pm$ 0.002   & 0.58 $\pm$ 0.002   & 0.58 $\pm$ 0.002  & 0.59 $\pm$ 0.004 \\
$F_{\mathrm{h}}$                  & 0.0                 & 0.0               & 0.0.               & 0.0.             & 0.0                & 0.0.                 & 0.0.             & 0.0 \\
$\chi$2 & 2.2 & 2.4 & 2.7 & 2.12 & 2.9 & 2.17 & 3.0 & 2.9 \\
\hline
\end{tabular}
}
\end{table*}

\clearpage
\newpage

\section{Azimuthally Modulated Ring Model \label{sec:ring_model}}

To analyze the flux  distribution at the inner disk of HD\,163296, the PIONIER and GRAVITY observables were fitted with an azimuthally modulated ring (hereafter called Ring model) based on the prescription proposed by \citet{Lazareff2017}. It is composed of an infinitesimal wire-frame ring with the following Fourier transform: 

\begin{equation}
R(u_r,v_r) = J_{0}(2\pi q \mathrm{a}_{r}) + \sum_{j=1}^{m}(-i)^j \rho_j \mathrm{cos}(j(\psi-\theta_j)) J_{j}(2\pi q \mathrm{a}_{r})\,,
\end{equation}

\noindent here, to account for the possible elongation of the ring, the original u-v coordinates were rotated and inclined in the following form:

\begin{align}
\label{eq:uv}
u_r &= u \mathrm{cos(PA)} + v\mathrm{cos(PA)} \\
v_r &= (-u \mathrm{sin(PA)} + v\mathrm{cos(PA)}) / \mathrm{cos(i)}
\end{align}

\noindent here, PA is the position angle (measured from East to North) of the semi-major axis of the ring, and cos(i) is the elongation factor. $q$ exp\,(i$\psi$) is the polar form of the spatial frequencies $u_r$-$v_r$. $\mathrm{a}_r$ is the ring angular radius. $\rho_j$\,exp(i$\theta_j$) is the polar representation of the modulation amplitude, $c_j$ + i$s_j$, applied to the ring profile. The index $j$ corresponds to the order of the modulation, in this case, we only tested models with $j$~=~1.  

To provide a width to the ring wire-frame, its profile is convolved with a Lorentzian kernel of the following form:


\begin{equation}
K(u_r,v_r) = \mathrm{exp}\left(\frac{-2\pi a_{k} q}{\mathrm{\sqrt{3}}}\right)\,,
\end{equation}
where $a_k$ is the Kernel angular radius. To avoid degeneracy in the fitting process, we fixed the quotient $a_k$ / $a_r$ = 0.3. The complete visibility expression is the following one:
\begin{equation}
V(u,v) = \frac{F_{\mathrm{s}} + F_{\mathrm{c}}  R(u_r, v_r) K(u_r, v_r)} {F_{\mathrm{s}} + F_{\mathrm{c}} + F_{\mathrm{h}}}\,,
\end{equation}

here, $F_{\mathrm{c}}, F_{\mathrm{s}}$ and  $F_{\mathrm{h}}$ are the flux contributions of the ring (denoted with the subscript "c"), point-like object (denoted with the subscript "s") and over-resolved component (denoted with the subscript "h"). We explicitly constrain the model to have $F_{\mathrm{s}} + F_{\mathrm{c}} + F_{\mathrm{h}} = 1$

For the minimization, we used the Gradient Descent Least-Squares algorithm implemented in \textit{lmfit}. The starting points of the parameters use values close to the position angle, inclination, size and flux ratios reported in \citet{Lazareff2017}. The $c_1$ and $s_1$ variables, which define the degree of asymmetry in the ring, were initially set to zero. Initial values were the same for all epochs and instruments. Each spectral channel was fitted independently. The 2013 datasets were combined into a single epoch. Since this corresponds to the same dataset used by \citet{Lazareff2017} and \citet{Kluska2020}, this serves us to compare our best-fit models directly with those previous estimates. For the PIONIER data sets taken in 2017, we only combined data taken with a maximum separation of three days and with a minimum of four snapshots. The 2018 and 2019 GRAVITY data are used as separate epochs. The best-fit parameters obtained for our model are reported in Table \ref{tab:param_ring_2013} and Table \ref{tab:param_ring_2018}. Figure \ref{fig:mean_gravity_images} displays a mean image of the best-fit model for each epoch. A comparison between the observables from the data and the ones recovered from our model can be consulted in Figures \ref{fig:ring_model2013v2} to \ref{fig:ring_model2019_cp} included in Appendix \ref{sec:observables_ring}.

\section{Results \label{sec:Results}}

\subsection{Constraints on the geometry of the target \label{sec:geometry}}
\begin{figure*}[thp]
\textbf{HD\,163296 - Best-fit azimuthally modulated ring model}\par\medskip
  \centering
  \includegraphics[width=18 cm]{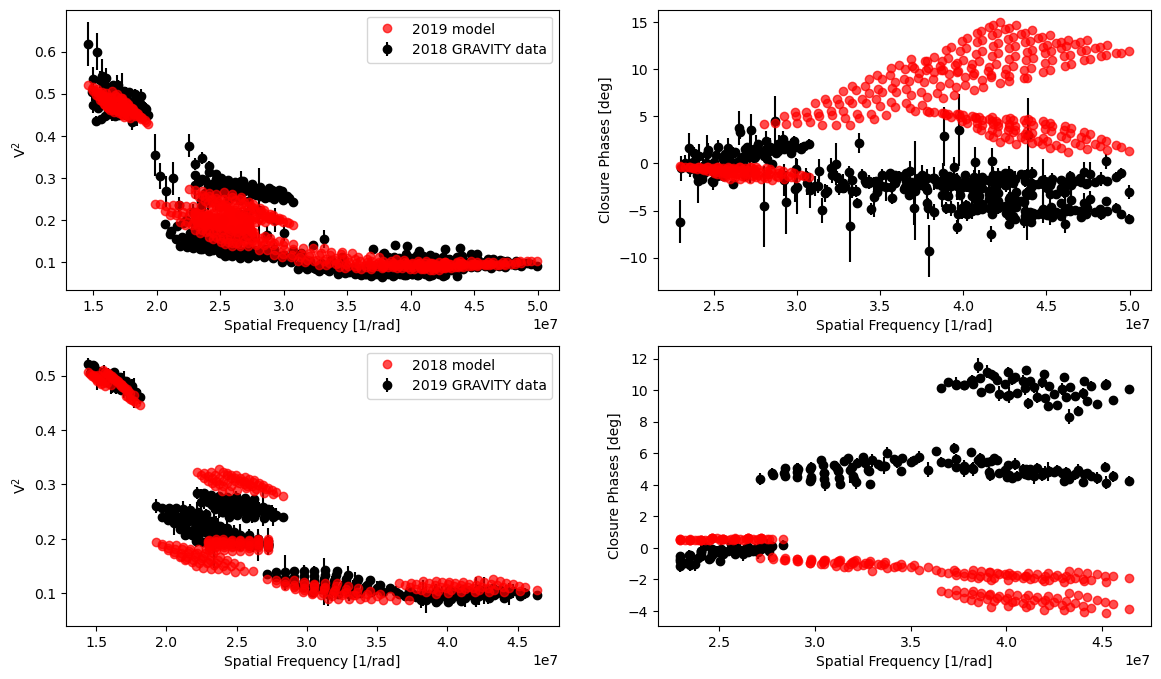}
\caption{The upper panels display the GRAVITY 2018 data with the observables extracted from the 2019 Ring model. The lower panels display the opposite case. Notice how the observables from one epoch are not reproduced by the model of the other one. This is more evident for the closure phases. Therefore, this test supports the presence of a variable-asymmetric structure at the inner disk in HD\,163296.}
\label{fig:obs_vs_ims}
\end{figure*}

Our parametric model reproduces the observables for all the epochs. The morphological characteristics of the source obtained with our best-fit model are the following ones:

\begin{enumerate}
\item \textit{Geometry:} The radii of the ring vary from 1.75 - 2.7 mas (0.175 - 0.27 au) for the 2013, 2018 and 2019 epochs, and of 0.65 - 1.2 mas (0.065 - 0.12 au) for both 2017 epochs. The 2013 PIONIER data helped us to compare our results with previous estimates in the literature. In this regard,  the estimated radius of the ring, $a_r$, is in agreement with the 1.81 mas reported by \citet{Lazareff2017} using the same dataset. We expect to observe a change in the radius of the ring between the $H$ and $K$ bands due to a radial gradient in the temperature profile of the dust. However, we observed that there is an important difference in the obtained values, even within the PIONIER epochs ($a_r^{2013, June/July} \sim$ 1.78 mas; $a_r^{2017,April} \sim$ 1.19 mas and; $a_r^{2017,Aug.} \sim$ 0.8 mas). 

  Our model predicts a projected disk semi-major axis position angle (North to East) of 130$^{\circ}$ - 140$^{\circ}$. This range of parameters is consistent with previous infrared interferometric findings, as well as with ALMA observations of the disk structure at large scale \citep[see e.g., ][ PA = 146$^{\circ}$ ]{Andrews2018}.

  The inclination angle of our model varies from  35$^{\circ}$ to 55$^{\circ}$ depending on the epoch. It is interesting to notice that, despite covering similar u-v frequencies, the GRAVITY data show a strong difference in the derived inclination angle between the 2018 ($i = 53.1\pm1.5$) and the 2019 ($i = 34.8\pm0.7$) epochs. We consider that this apparent difference in the inclination angle derived by our model is caused by the asymmetric and variable structure of the ring obtained from the closure phase information.

  \citet{Muro-Arena_2018} suggest a small misalignment from +1$^{\circ}$ to +3$^{\circ}$ between the spatially unresolved inner disk (with a radius between 4 and 10 mas) and the outer structure of the disk (which has an inclination of $i = 45^{\circ}$). The models presented by those authors support this claim in order to explain the outer shadow casts observed in the disk. It is interesting to notice that the inner inclination angle derived by \citet{Muro-Arena_2018}, based on SPHERE/IRDIS 2016 data, is consistent with the value derived with our 2017 April data ($i = 48.3\pm0.4$).

\item \textit{Flux Distribution:} F$_*$ contributes between 33\% and 45\% of the total flux, depending on the epoch, while F$_{\mathrm{c}}$ contributes between 40\% to 60\%, and F$_{\mathrm{h}}$ contributes between 0\%  to 20\%. The observed changes in the flux contributions of the PIONIER data could be, partly, explained by the sampling of different u-v frequencies between the different epochs. We notice that these large variations (more than 3$\sigma$) are also present in the GRAVITY data. Since our GRAVITY epochs have coincident baselines, the observed variations in the GRAVITY data cannot be related to flux filtering due to the different u-v spacing. In contrast, our findings suggest that there is an intrinsic flux variability associated with changes in the morphology of the source. This is in direct line with the results from \citet{Kobus_2020} who interpret the variability of the squared visibilities for similar baselines at different epochs as changes in the structure of the target.

\item \textit{Asymmetry:} We can appreciate that the loci of the brightest side of the Ring model is changing, which strongly supports the variability of the inner ring structure. Notice that the values of the coefficients $c_1$ and $s_1$, which traces the modulation of the ring, are changing between epochs. Observing these changes is of particular interest for the GRAVITY data, since the 2018 and 2019 epochs sample quasi-coincident spatial frequencies (in length and position angle). Figure \ref{fig:obs_vs_ims} displays the GRAVITY observables for the 2018 epoch over-plotted with the observables extracted from the best-fit models obtained from the 2019 epoch, together with the opposite case. It can be noticed that observables extracted from the model in one epoch are not able to reproduce the observables of the other one, despite having quasi-coincident u-v frequencies. This demonstrates changes in the morphology of the object in, at least, temporal scales of 1 year for the GRAVITY data. These time-scales are restricted by the temporal sampling of our data. However, we cannot exclude changes in the structure at shorter time-scales and/or the possibility of being observing different bright structures instead of a persistent single one. 

  The presence of an asymmetry and hints in the variability were suggested before by several authors in the literature. The two most recent ones are \citet{Kobus_2020} and \citet{Varga_2020}. The first work uses archival PIONIER data and parametric models applied to the squared visibilities in order to derive morphological changes in the structure of the source. These authors infer that the source must have symmetrical variations. Nevertheless, they do not include models to  fit simultaneously closure phases  and squared visibilities. Therefore, they cannot prove the asymmetric nature of the source. The second work uses MATISSE data and parametric models to constrain the source in $L$-band. However, this study includes a limited number of large baselines, where the asymmetry is detected. Therefore, the authors cannot confirm the variability of the target solely with the MATISSE data. Moreover their work just traces the asymmetric structure of the extended emission at one position angle. In contrast, our findings go one step forward to confirm the asymmetry of the source, its variable nature, and persistence (on a temporal baseline of 7 years).
  
\end{enumerate}

\begin{table*}[htp]
\caption[]{HD\,163296 - Temperature of the asymmetry}
\label{tab:temperature}
\centering
\begin{tabular}{l c c c c } 
\hline \hline
Epoch & Filter & $\Delta$RA$_{\mathrm{peak}}$ [mas] & $\Delta$Dec$_{\mathrm{peak}}$ [mas] & Temperature [K] \\
\hline
2013 (June/July) & H-band & -2.26 $\pm$ 0.05  & 0.26 $\pm$ 0.12 & 1300 $\pm$ 90 \\
2017 (April) &  H-band  & -0.1 $\pm$ 0.71   & -1.2 $\pm$ 0.41 & 1620 $\pm$ 240 \\
2017 (Aug.) &  H-band  & 0.8 $\pm$ 0.41     & -1.1 $\pm$ 0.11 & 1580 $\pm$ 140 \\
2018 (June) &  K-band  &  0.01 $\pm$ 0.15   & 2.3  $\pm$ 0.17 & 1330 $\pm$ 90 \\
2019 (June) &  K-band & -2.7 $\pm$ 0.08 & -0.7 $\pm$ 0.24 & 1240 $\pm$ 80 \\
\hline
\end{tabular}
\end{table*}

\subsection{Limitations of the azimuthally modulated ring model \label{sec:limitations}}

While the model presented in Sec. \ref{sec:ring_model} reproduces the different trends observed in the data, we have identified the limitations discussed below:

\textit{Comparison with other parametric models:} Our azimuthally modulated ring model uses a valid prescription to describe the morphology of the target at the spatial scales traced with PIONIER and GRAVITY. Similar ring models have been presented in the literature \citep{Lazareff2017, Kluska2020, Perraut2019, Varga_2020} with good results to reproduce the morphology of the target. Nevertheless, we consider that our model presents several limitations that need to be taken into account when interpreting the morphology of the target. These limitations are related to: (i) a limited u-v coverage, particularly for the GRAVITY data; (ii) the lack of baselines larger than 100 m to clearly sample the visibility trend after the first rebound in the Fourier transform of the ring and; (iii) the lack of similar baselines and wavelength sampling on all epochs.

We complemented our analysis by exploring another parametric model based on an off-centered Gaussian plus a central point-like source and an over-resolved component (hereafter called Gaussian model). A detailed description of this Gaussian model is included in Appendix \ref{sec:gaussian_model}. By comparing the two models, we confirmed the following points:

\begin{itemize}
\item Both models are able to reproduce the trend observed in the squared visibilities and closure phases with similar degree of accuracy (see the reduced $\chi^2$ values reported in Tables \ref{tab:param_ring_2013} - \ref{tab:param_ring_2018} and Tables \ref{tab:param_model_2013} - \ref{tab:param_model_2019} in Appendix \ref{sec:gaussian_model}). Therefore, both are equally valid solutions to describe the morphology of HD\,163296. Similar inclination and position angles are found between the Ring model and the Gaussian one. The Gaussian model also shows the difference in the inclination angle of 50$^{\circ}$ and 30$^{\circ}$  between the 2018 and 2019 GRAVITY data.

However, we find a larger discrepancy in the contribution of $F_s$ to the total flux compared with the Ring model. While the Ring model predicts values of $F_s$ between 33\% and 45\% of the total flux, the Gaussian model predicts values between 20\% and 30\%. This discrepancy can be explained by the fact that the Fourier transform of a Gaussian is a continuous and monotonically asymptotes to the F$_s$ value as spatial frequencies increase. Therefore, an additional contribution from the Gaussian is always present on top of the flux contribution of the central source to reproduce the visibility trends. This fact compensates the lower percentage of $F_s$ obtained with the best-fit Gaussian model compared with the Ring one.

The off-centered Gaussian model also reproduces the asymmetries of the source structure. In this model, the direction of the peak's displacement (asymmetry vs central source) is in agreement with the brightest side of the Ring model for each different epoch. Nevertheless, the amplitude of the displacements of the Gaussian component is one order of magnitude smaller than the position of the brightest part of the Ring. This is because the displacements of the Gaussian component are tracing the flux-centroid position of the extended morphology and not the position of the asymmetry in the structure. Therefore, caution must be taken when comparing those values with the position of the brightest side of the Ring model.

\item Given the limitations of our Ring model,  it is interesting to compare it with previous formulations in the literature, which use a Gaussian to describe the extended morphology of the target. For example, \citet{Setterholm_2018} use radially symmetric models: a Gaussian, a uniform disk and an infinitesimal ring. All of them include a centered point-like source The data used by those authors consisted in a combination of different instruments from the VLTI and CHARA. Their u-v sampling include baselines up to 350 meters and their models were applied to the $H$ and $K$ bands. Their results suggest that an on-axis elongated Gaussian and a point-like source reproduce better the visibility function of the target. In particular, those authors found problems to reproduce the short angular scales (the ones traced with baselines < 50 m.) with the uniform disk and infinitesimal ring models. However, we do not find a similar problem to reproduce the visibilities at short spatial scales using our Ring model.

One important difference between the ring model presented by \citet{Setterholm_2018} and ours is that those authors use an infinitesimal ring, which produces more pronounced rebounds in the visibilities after the first minimum. In our case, the wire-frame of our azimuthally modulated ring is convolved with a Lorentzian kernel which smooths the profile of the ring and produces flatter rebounds after the first minimum in the visibility trend. Additionally, the uncertainties in the $K-$band data presented by those authors have values as large as $\sigma_{V^2} \sim$ 0.1 for baselines above 100 m. This makes difficult to ensure the monotonic decrement of the visibility.

Furthermore, visibilities in $H-$band appear to have a small rebound for the largest baselines above 200 m. Nevertheless, the $H-$band data lack of intermediate baselines. This limitation does not provide us with more robust estimates of the visibility profile. The differences in the disk' sizes reported in Sec. \ref{sec:geometry} support that the inner structure of the disk does not have a sharp edge. In contrast, it appears that the inner disk is smooth, being the best-fit rings of our model the emission traced by the different interferometric arrays used (see also Sec.\,\ref{sec:nature_asymmetry}). 

\textit{Comparison with reconstructed images:} Our geometrical models reproduce the observed profiles in our data. However, those geometrical models can only explore a limited degree of asymmetries in the data. In order to better explore the asymmetries traced by the closure phases, we performed image reconstruction on our datasets. The complete imaging process is described in Appendix \ref{sec:app_imaging}. The regularized minimization used for image reconstruction is able to trace more complex asymmetric structures in the data. It is true that imaging is more suitable for rich u-v coverages, therefore, our limited data constrain the quality of the reconstructions. Still, by comparing independent images from the data and reconstructed ones from our parametric models is possible to improve our knowledge about the asymmetric morphology of HD\,163296.

From the best images obtained, we could not favor the Ring nor the Gaussian model as the best one to describe the target. This is because reconstructed images obtained from their simulated observables are quite similar to the recovered images obtained  using our data. However, we observed that, for some spectral channels, there are residuals as large as 20\% of the peak in the images. This is similar for both the Gaussian and Ring models. Therefore, this supports that the degree of asymmetry of the source traced by our parametric-models is underestimated and the morphology of the target is more complex than what we map with the geometrical models.

\end{itemize}

\section{Discussion \label{sec:Discussion}}

\subsection{Temperature of the asymmetry:}

The main result of our analysis is the confirmation of the asymmetric and variable structure of the inner disk morphology of HD\,163296. We expect to have an optically thick inner disk. Therefore, estimating the temperature of the dust (T$_d$) at the positions of the asymmetry cannot be obtained directly with its surface brightness. However, we can compute a rough estimate of T$_d$ by assuming dust grains of given sizes directly heated by the UV radiation of the central source, which are located at a distance $r$ from the star at the position of the brightest point in our ring model. For this purpose, we use the following expression \citep{vanBuren_1988}:

\begin{equation}
  \mathrm{T_d} = 27\mathrm{a}^{-1/6}_{\mu m}\, \mathrm{L}_{*,38}^{1/6}\, \mathrm{r}^{-1/3}_{pc}\,\,\mathrm{K}\,
\end{equation}


here,  L$_{*,38}$ is the UV luminosity of the star in units of 10$^{38}$ erg\,s$^{-1}$; r$_{pc}$ is the distance to the dust from the star in units of parsecs; and ,$a$, is the size of the dust particle in microns. T$_d$ was computed considering a power-law dust size distribution \citet[][]{Mathis_1977}, for dust sizes ranging from 10$^{-2}$ to 1~$\mu$m. To calculate the distribution of temperatures per epoch, we extracted 10$^{4}$ different samples of the peak's positions, dust grain sizes, and we set L$_{*,38}$~=~6.38$\times$10$^{34}$~erg\,s$^{-1}$ \citep{Acke_2004}. The deprojected peak's positions were obtained from the Ring model, assuming an inclination i~=~40$^{\circ}$. This produces the temperatures reported in Table\,\ref{tab:temperature}. These values ( 1240 K < T$_d$ < 1600 K) are in agreement with  temperatures between the sublimation point of the silicate (T$_s$ = 1500 K) and carbon (T$_c$ = 1800 K) dust grains. However, we should notice that the reported range of temperatures only traces the material observed with the different interferometric datasets. Therefore, those temperature values do not necessary correspond to the upper temperature limit of the most heated dust in the disk.

Previous near-IR interferometric studies \citep{Benisty_2010, Setterholm_2018} suggested the possibility of having refractory dust grains (which survive temperatures above T > 2000 K). At mid-infrared wavelengths, measurements obtained with MIDI-VLTI and reported by \citet{van_Boekel_2004} found a considerable larger fraction of cristallinity within the central 20 mas (2 au) in HD\,163296, compared with the outer  20-200 mas (2-20 au) of the disk (40\% $\pm$ 20\% vs 15\% $\pm$ 10\%). Similarly, the ring models reported by \citet{Varga_2020}  support that around 20\% of the surface brightness near the star comes from a region where small micron-sized standard dust grains cannot survive. Hence, those authors, also, suggest the presence of refractory grains with small cooling efficiencies ($\epsilon \sim$ 0.1 = 0.18) which survive temperatures above 2300 K. These constraints and our derived T$_d$ values support the presence of a mixing dust species or refractory dust grains as responsible for the thermal emission of the variable disk.

\begin{figure}[thp]
\textbf{HD\,163296 - positions of the emission's peak asymmetry in the Ring model}\par\medskip
  \centering
  \includegraphics[width=7 cm]{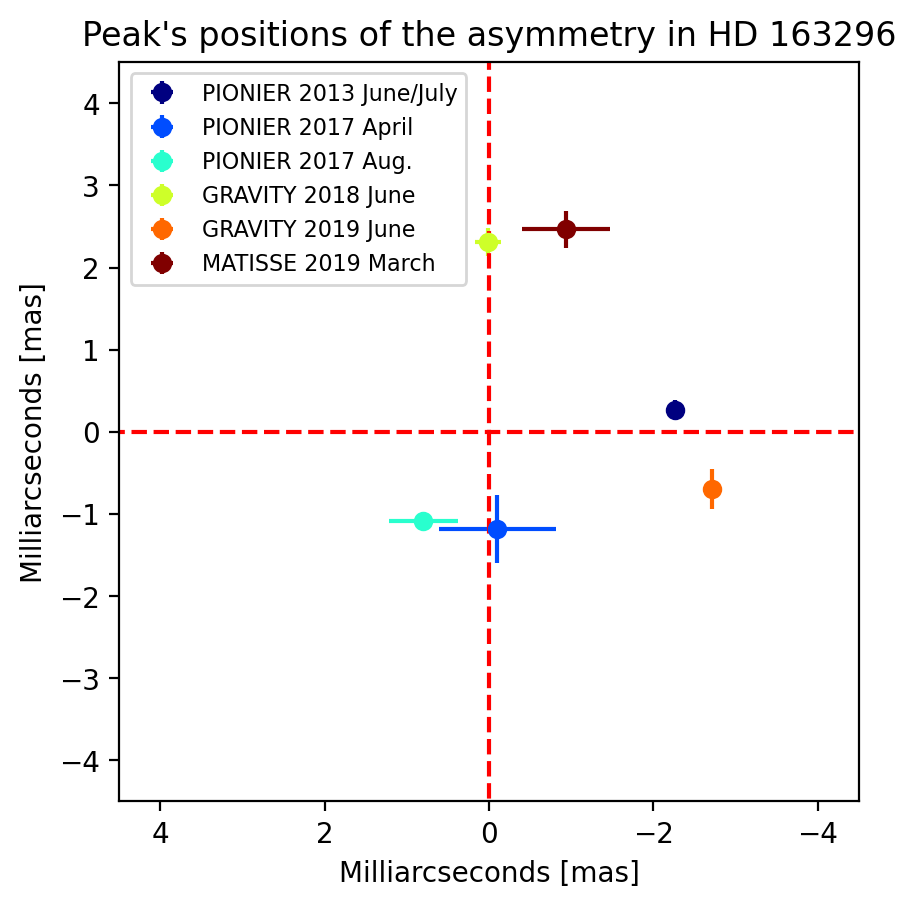}
\caption{ The plot shows the positions of the emission's peak extracted from our Ring model and complemented with the position of the peak obtained from the MATISSE data and the ring model presented by \citet{Varga_2020}.}
\label{fig:positions}
\end{figure}

\subsection{On the nature of the asymmetry \label{sec:nature_asymmetry}}

The derived changes in the size of the fitted ring cannot be explained solely by the temperature gradient since the different epochs are sampled at the same wavelength. Therefore, they can only be explained by the changes in the effective resolution between the different epochs (see Fig.\,\ref{fig:uv_planes}). With larger baselines, the 2017 configurations filter out most of the extended emission of the source and, therefore, a more compact object is observed by the interferometer, being the 2017 August epoch the one with the smallest estimate of the ring size. These results indicate that our observations do not trace a sharp edge of the inner ring structure. In contrast, they support the existence of a more smooth inner morphology of the disk. 

From our best-fit models, the peak of the emission in the ring changes for each one of the different epochs. Figure \ref{fig:positions} displays the projected positions of the peak in the ring emission for PIONIER and GRAVITY, complemented with the peak's position from the ring model applied to the MATISSE observations described in \citet{Varga_2020}.  Unfortunately, we have too limited amount of data to clearly trace the orbital motion of the material and to test the presence of a single persistent structure, instead of several different ones, over the 7 years spanning our observations. Furthermore, due to the change in angular resolution, the apparent size of the ring changes by a factor of two between the PIONIER and the GRAVITY models (see Fig.\,\ref{fig:mean_gravity_images}). Our limited resolution, does not allow us to clearly resolve the asymmetry, therefore, we cannot determine whether its forming material is distributed on a well-localized structure or, if it is more extended over several angular scales. In this section, we discuss several physical scenarios to explain the origin of the observed asymmetry.

Due to the change in distance and position angle of the asymmetry across the different epochs and instruments, we discard the possibility of being observing an illumination effect due to a fixed inclination of the disk. A possible cause of such an asymmetry is that the inner disk is warped and, therefore, casts different shadows on the outer disk. Very recently, \citet{Kraus_2020} discovered a highly misaligned and warped disk around GW Orionis. These authors propose disk tearing \citep{Facchini_2013} as the hydrodynamic effect that causes the inner disk to change its orientation and precession. However, this mechanism is only possible if there is at least a stellar binary as central engine for the system. In the case of HD\,163296, there is no evidence of a secondary stellar companion. Furthermore, the inner disk shows small or no precession, as indirectly seen from the jet/counter jet opening angle \citep[$\sim$2$^{\circ}$;][]{Wassell_2006}.


Another possibility to explain the nature of the asymmetry is the presence of a local perturbation on the disk material. An interesting hypothesis is the presence of a pressure bump produced by a vortex originated from an unseen planetary or dwarf companion. Until now, efficient dust traps produced by an anticyclonic vortex have been presented as plausible explanations for large (mm) dust grains to be trapped in an azimuthal direction on the disk \citep{vanDerMarel_2013, vanDerMarel_2018, Pineda_2019}. These dust traps tend to create arc-like features similar to the observed ones in our ring-based model. Additionally, magneto-hydrodynamical simulations conducted by \citet{Flock2017} show that a local pressure maximum inside the disk's dead zone favors the creation of vortex, which can cast non-axisymmetric shadows on the outer disk.

More recently, the simulations performed by \citet{Varga_2020} suggest that a large scale vortex produced by a Rossby wave instability could be the cause of the asymmetry in HD\,163296. It is important to mention that density enhancement is not enough to create a change in the brightness distribution of the ring profile. This is because the emission of the ring is expected to be optically thick. In the scenario proposed by \citet{Varga_2020}, the large scale vortex favors the production of small dust grains, which modify the local temperature profile of the disk and, therefore, produce an increment in the emission at the position of the vortex. 

Finally, the observed asymmetry could be explained by an in-homogeneous distribution of dust above the mid-plane of the disk. This idea is supported by recent HST data of the outer disk structure. \citet{Rich_2019, Rich_2020} reported strong surface brightness variations at scales larger than 660 mas (66 au) on time scales lower than 3 months. These results suggest that the origin of the moving shadows is material located at distances smaller than 5 mas (0.5 au) from the central star. To produce shadows in the disk at large scales, the material must reside at 0.8 mas (0.08 au) above the mid-plane of the disk, assuming coplanarity of the inner and outer disk. These authors also reported the presence of two dipper events in 2018, probably caused by variations of the scale height of the inner rim. The material must reside at the inner 4.1 mas (0.41 au) and at a scale height above 3.7 mas (0.37 au), suggesting the presence of a dusty wind.

A theoretical dusty-wind model that lifts material above the mid plane has been proposed by \citet{Bans_2012, Ellerbroek_2014}. Those authors also suggested that such a model is an important candidate for the origin of the strong outflows, like the one present in HD\,163296. This would support the existence of material ejected above the mid-plane of the disk, not homogeneously distributed, which might be linked to the variable structure that we observe. To conclude which of the aforementioned scenarios is more plausible to explain the asymmetry in HD\,163296, more observations (to improve considerably our u-v plane) are required in addition with dedicated simulations of the object.

\section{Conclusions \label{sec:Conclusions}}

This work presents new near-infrared interferometric observations of HD\,163296 taken with GRAVITY, complemented with archival PIONIER data. Our multi-epoch campaign allows us to characterize the asymmetric and variable inner structure of the target. For this purpose, we used a parametric model of an azimuthally modulated ring. This model reproduces the squared visibilities and closure phases of each one of the epochs analyzed. To test the limitations of our model, we also fitted the data with an off-centered Gaussian model and, we conducted image reconstruction. These additional model and images confirmed the asymmetry of the inner disk and its variability. The inclination and position angle of the disk found with our parametric models are in agreement with previous estimates. However, the changes in the size of the ring across the different epochs make us support that the disk does not have a sharp inner edge but a smooth brightness profile.

Due to the variability of the disk morphology, we hypothesize that the nature of the asymmetry is not caused by an illumination effect. More plausible explanations include the presence of a local perturbation, like a vortex, or the presence of ejected dust above the mid-plane of the disk. Our estimation of the temperature of the asymmetry favors the existence of a mixed population of carbon and silicate dust grains or, as previously suggested, the presence of refractory dust grains. New data taken with MATISSE add further evidence for the presence of a non-centrosymmetric structure over different angular scales across the $H$, $K$ and $L$ bands. To fully determine the nature of a such structure, it is necessary to combine several interferometric observations with different baselines and wavelengths. Due to the  high variability of the source, it is critical to obtain data over short time-scales (less than a month) to properly combine them and being able to perform image reconstruction and more sophisticated parametric (radiative transfer) models to unveil the nature of the asymmetry.

 
\begin{acknowledgements}
The authors thank the anonymous referee for his/her valuable comments. JSB acknowledges the full support from the UNAM PAPIIT project IA 101220. This research has made use of the Jean-Marie Mariotti Center OiDB service available at http://oidb.jmmc.fr. ACG acknowledges funding from the European Research Council under Advanced Grant No. 743029, Ejection, Accretion Structures in YSOs (EASY). A.A., M.F. and P.G. were supported by Funda\c{c}\~{a}o para a Ci\^{e}ncia e a Tecnologia, with grants reference UIDB/00099/2020 and SFRH/BSAB/142940/2018. RGL acknowledges support from Science Foundation Ireland under Grant No. 18/SIRG/5597.
\end{acknowledgements}

%
%
\bibliographystyle{aa} 
\bibliography{ref.bib}

\clearpage
\newpage

\begin{appendix}
  \section{Interferometric Observations (Log and u-v planes)}
  
  \begin{table*}[htp]
\caption[]{HD\,163296 Observational Log}
\label{tab:observations}
\centering
\begin{tabular}{l c  c c c c} 
\hline \hline
\multicolumn{6}{c}{GRAVITY Observations$^3$} \\
Date (dd/mm/yyyy) & Array & No. of Snapshots & Average Seeing$^1$ & Max. Resolution$^2$ & Min. Resolution \\
\hline
07-07-2018 & D0-G2-J3-K0 & 12  & 0.63$\pm$0.13 & 2.25 mas & 6.50 mas\\
14-07-2019 & D0-G2-J3-K0 & 7 & 1.22$\pm$0.05 & 2.42 mas & 6.58 mas  \\
\hline
\multicolumn{6}{c}{PIONIER Observations} \\
Date (dd/mm/yyyy) & Array & No. of Snapshots & Average Seeing & Max. Resolution & Min. Resolution \\
\hline
14-06-2013 & D0-G1-H0-I1 & 1 & 0.96 & 2.38 mas & 4.78 mas \\
02-07-2013 & A1-B2-C1-D0 & 3 & 1.07 $\pm$ 0.24 & 4.91 mas & 16.01 mas \\
22-04-2017 & D0-G2-J3-K0 & 4 & 0.44$\pm$0.04 &  1.63 mas & 4.28 mas \\
29-08-2017 & A0-G1-J2-J3 & 4 & 0.70$\pm$0.25 &  1.28 mas & 3.58 mas \\
30-08-2017 & A0-G1-J2-J3 & 1 & 0.63 &  1.28 mas & 3.17 mas \\
\hline
\end{tabular}
\begin{list} {}{} \itemsep1pt \parskip0pt \parsep0pt \footnotesize
      \small
      \item $^1$ Average seeing and its standard deviation computed over the total of snapshots per configuration and day
      \item $^2$ Resolution is defined as $\theta$ = $\lambda$ / (2B$_{max}$). For the values reported, we used $\lambda_{0}$ = 2.2 $\mu$m, and $\lambda_{0}$ = 1.65 $\mu$m as central wavelengths for GRAVITY and PIONIER, respectively 
    \end{list}
    
  \end{table*}
  
\begin{figure*}
\textbf{HD\,163296 - (u,v) planes}\par\medskip
  \centering
  \includegraphics[width=16 cm]{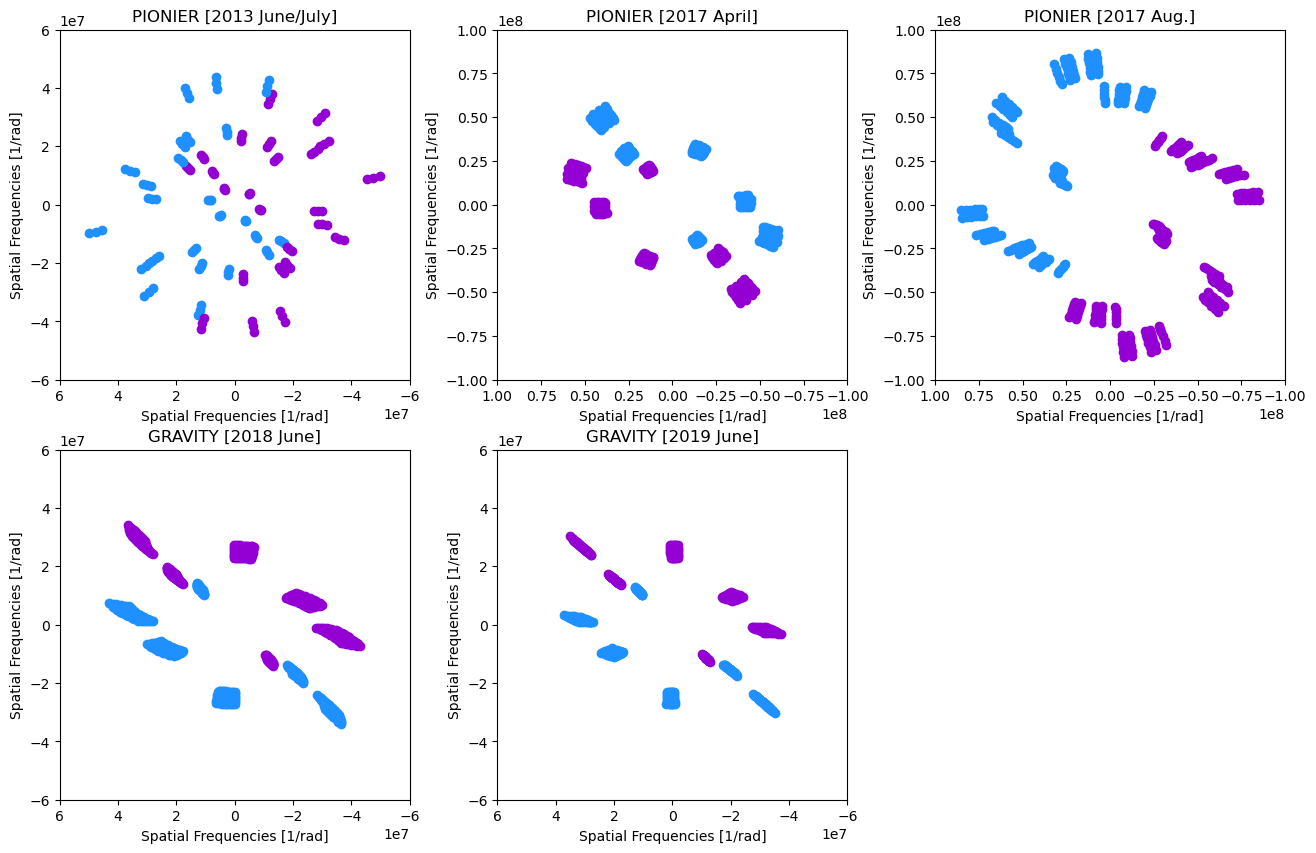}
\caption{u-v sampling of the different instruments and epochs used for this study. Blue dots indicate the spatial frequencies sampled while the violet ones indicate their complex conjugate.}
\label{fig:uv_planes}
\end{figure*}

\clearpage
\newpage

 \section{Interferometric observables and best-fit azimuthally modulated ring model \label{sec:observables_ring}}

\begin{figure*}[htp]
\textbf{HD\,163296 - Ring Model (2013 June/July, V$^2$)}\par\medskip
  \centering
  \includegraphics[width=13 cm, height=5 cm]{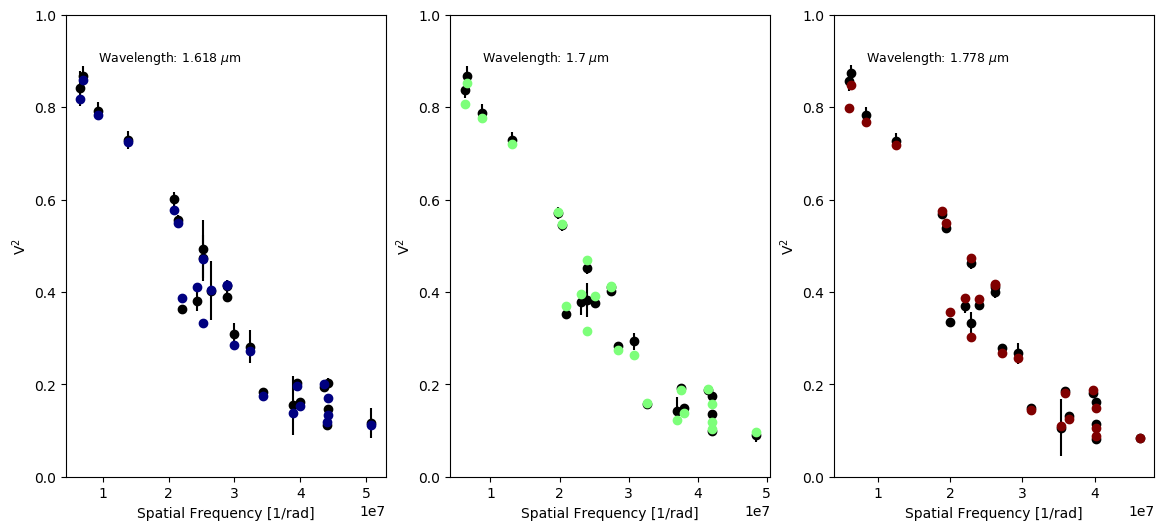}
\caption{Observations vs Ring model for the 2013 run. Panels display the squared visibilities (black dots) from the data versus spatial frequency. Each panel corresponds to a different wavelength (see labels on the panels). The synthetic observables obtained from the best-fit Ring model are over-plotted with colored dots in the different panels.}
\label{fig:ring_model2013v2}
\end{figure*}

\begin{figure*}[htp]
\textbf{HD\,163296 - Ring Model (2013 June/July, CPs)}\par\medskip
  \centering
  \includegraphics[width=13 cm , height=5 cm]{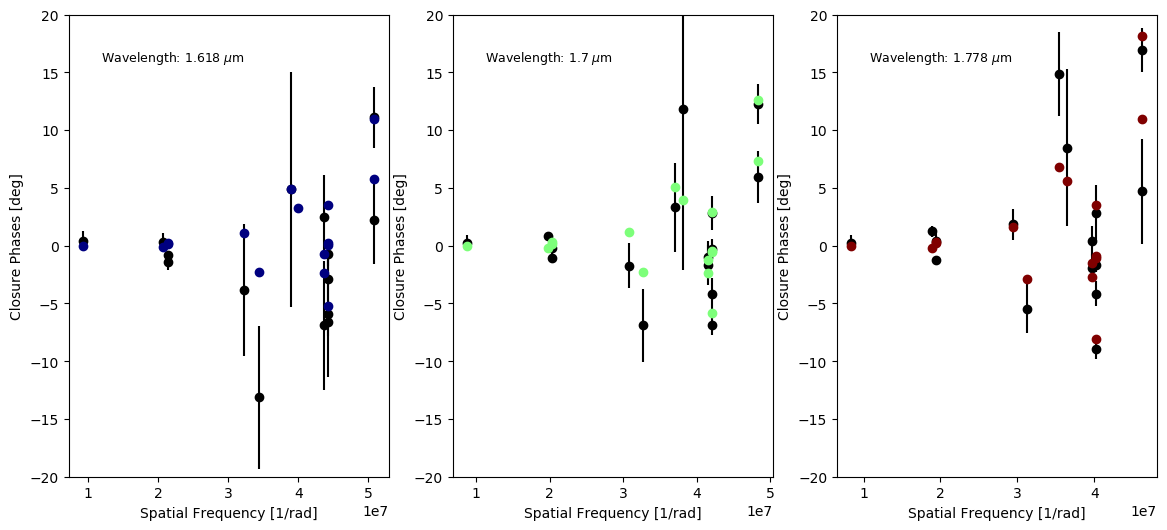}
\caption{Observations vs Ring model for the 2013 run. Panels display the closure phase (black dots) from the data versus spatial frequency. Each panel corresponds to a different wavelength (see labels on the panels). The synthetic observables obtained from the best-fit Ring model are over-plotted with different colors in the panels.}
\label{fig:ring_model2013cp}
\end{figure*}

\begin{figure*}[htp]
\textbf{HD\,163296 - Ring Model (2017 April, V$^2$)}\par\medskip
  \centering
  \includegraphics[width=\linewidth]{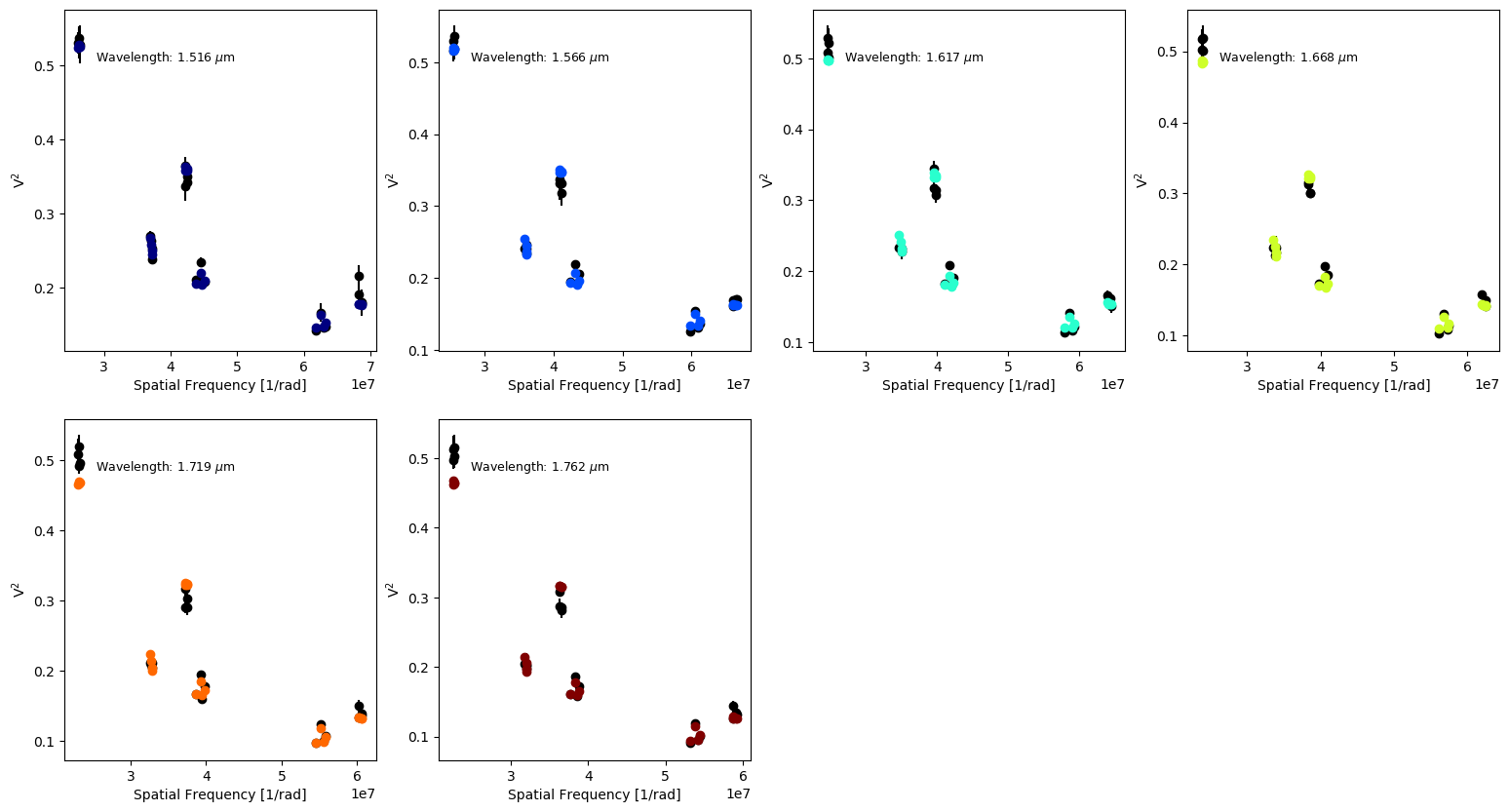}
\caption{Observations vs Ring model for the 2017 April run. Panels are plotted as in Fig. \ref{fig:ring_model2013v2}}
\label{fig:ring_model2017_1_v2}
\end{figure*}

\begin{figure*}[htp]
\textbf{HD\,163296 - Ring Model (2017 April, CPs)}\par\medskip
  \centering
  \includegraphics[width=\linewidth]{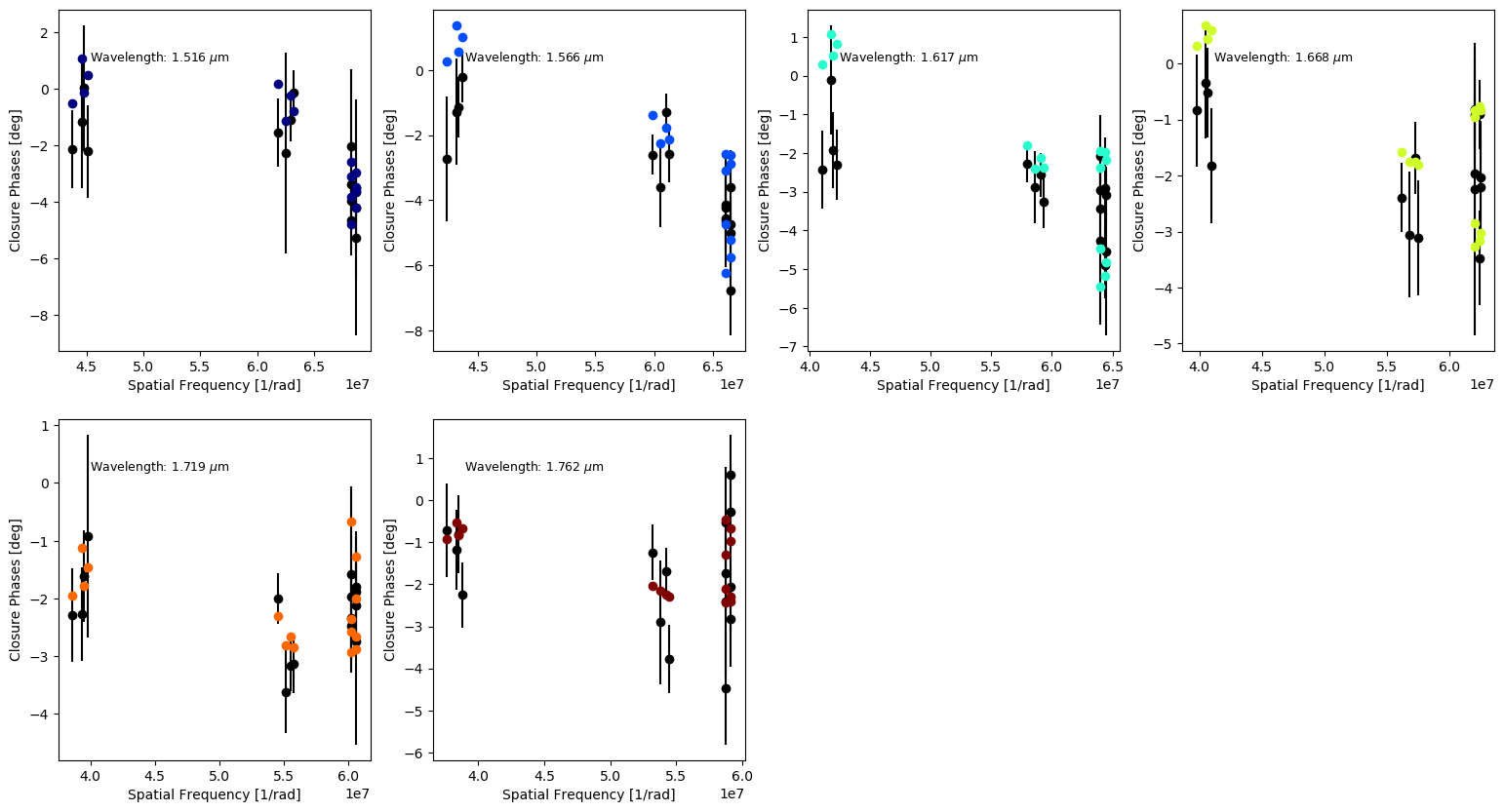}
\caption{Observations vs Ring model for the 2017 April run. Panels are plotted as in Fig. \ref{fig:ring_model2013cp}}
\label{fig:ring_model2017_1_cp}
\end{figure*}

\begin{figure*}[htp]
\textbf{HD\,163296 - Ring Model (2017 Aug., V$^2$)}\par\medskip
  \centering
  \includegraphics[width=\linewidth]{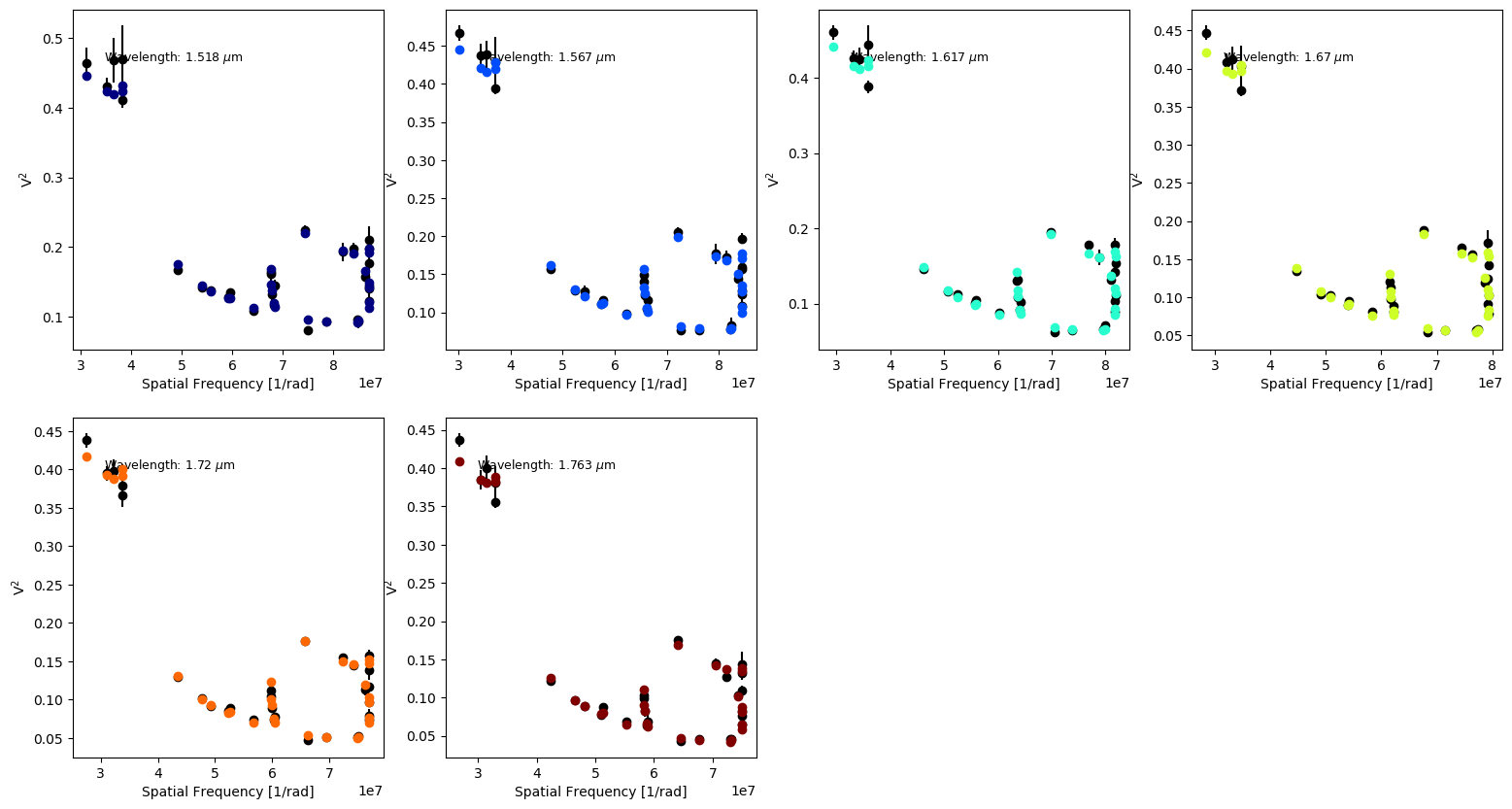}
\caption{Observations vs Ring model for the 2017 Aug. run. Panels are plotted as in Fig. \ref{fig:ring_model2013v2}}
\label{fig:ring_model2017_2_v2}
\end{figure*}

\begin{figure*}[htp]
\textbf{HD\,163296 - Ring Model (2017 Aug., CPs)}\par\medskip
  \centering
  \includegraphics[width=\linewidth]{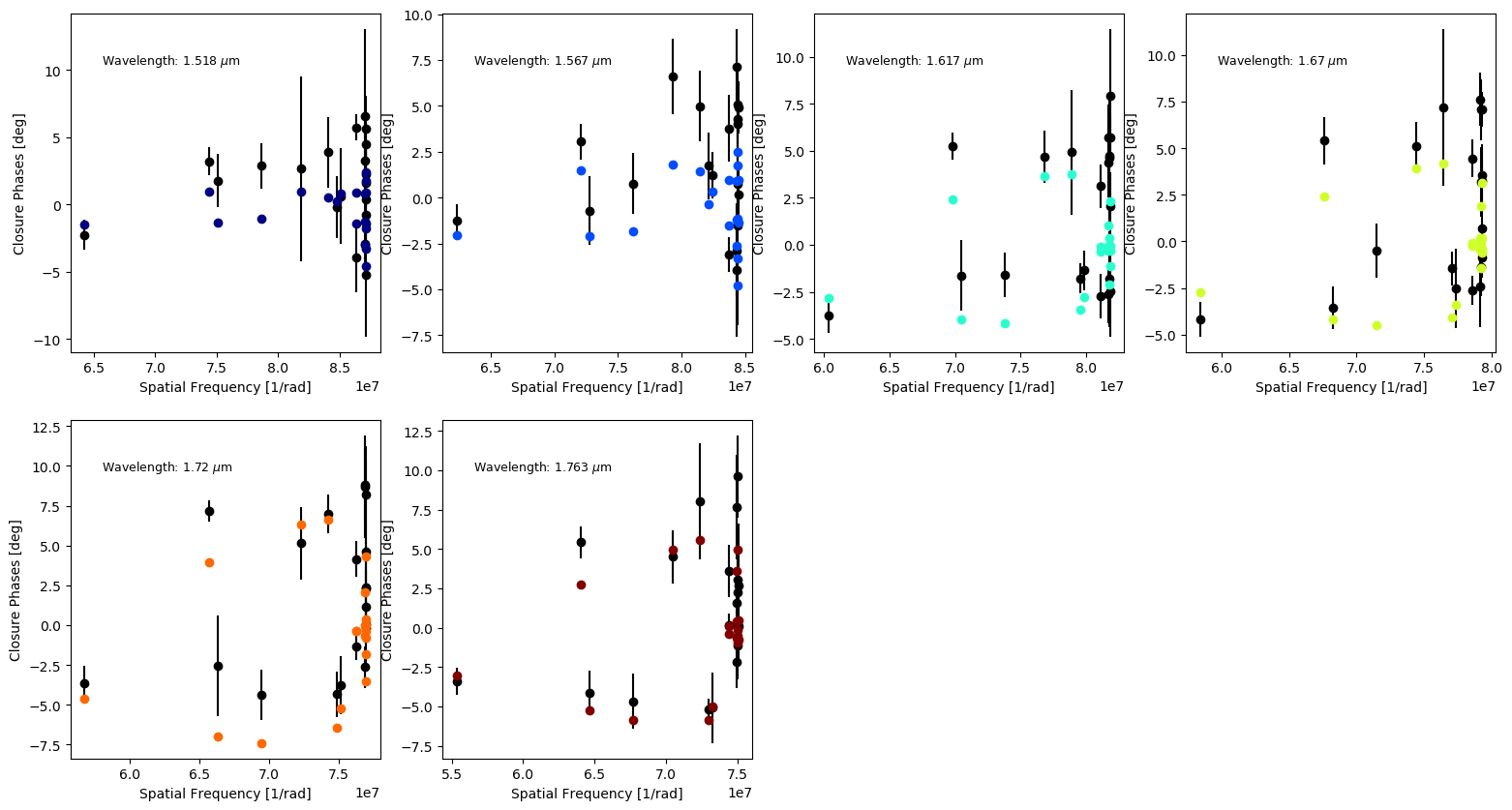}
\caption{Observations vs Ring model for the 2017 Aug. run. Panels are plotted as in Fig. \ref{fig:ring_model2013cp}}
\label{fig:ring_model2017_2_cp}
\end{figure*}

\begin{figure*}[htp]
\textbf{HD\,163296 - Ring Model (2018 June, V$^2$)}\par\medskip
  \centering
  \includegraphics[width=\linewidth]{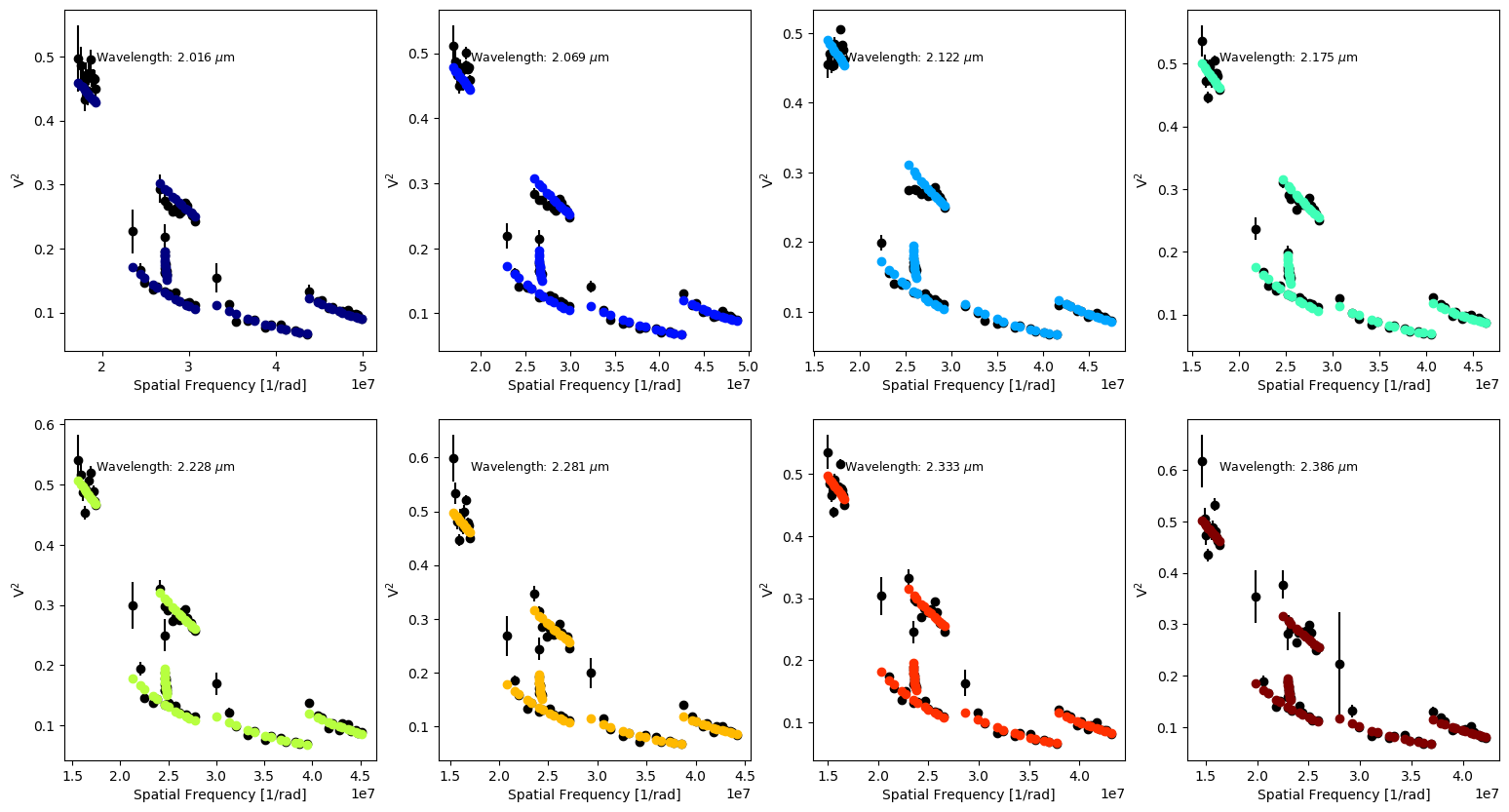}
\caption{Observations vs Ring model for the 2018 run. Panels are plotted as in Fig. \ref{fig:ring_model2013v2}.}
\label{fig:ring_model2018_v2}
\end{figure*}

\begin{figure*}[htp]
\textbf{HD\,163296 - Ring Model (2018 June, CPs)}\par\medskip
  \centering
  \includegraphics[width=\linewidth]{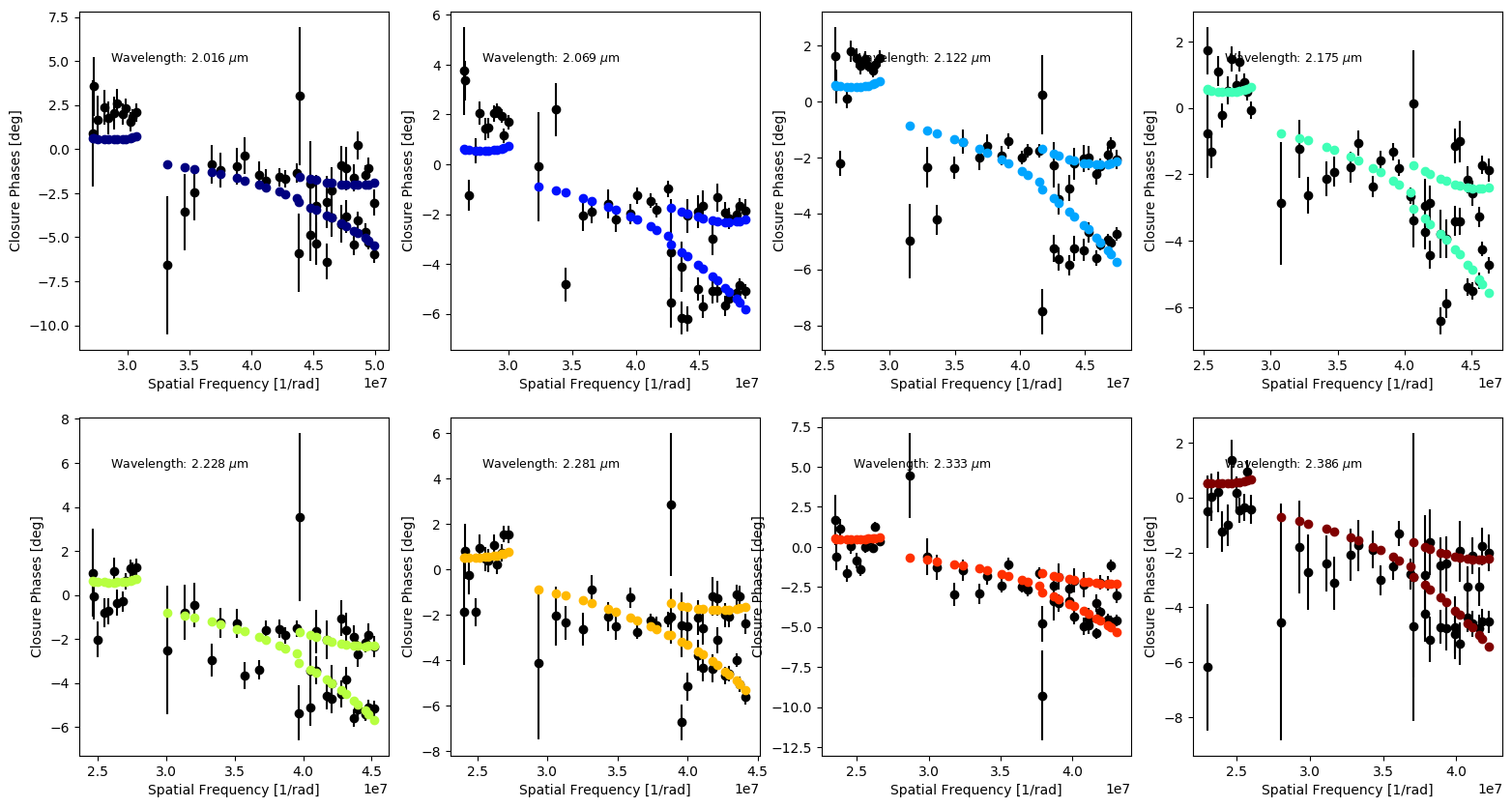}
\caption{Observations vs Ring model for the 2018 run. Panels are plotted as in Fig. \ref{fig:ring_model2013cp}.}
\label{fig:ring_model2018_cp}
\end{figure*}

\begin{figure*}[htp]
\textbf{HD\,163296 - Ring Model (2019 June, V$^2$)}\par\medskip
  \centering
  \includegraphics[width=\linewidth]{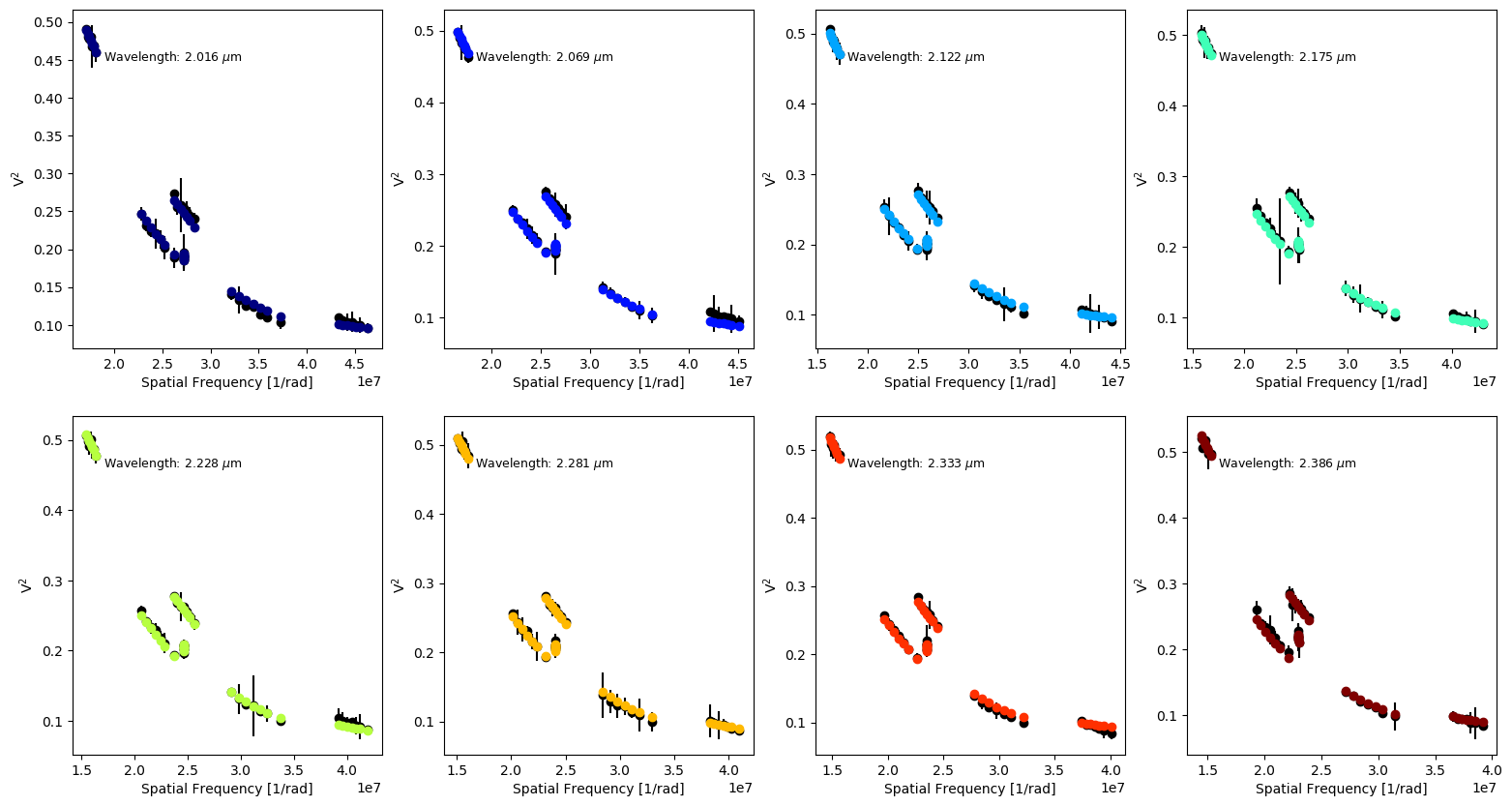}
\caption{Observations vs Ring model for the 2019 run. Panels are plotted as in Fig. \ref{fig:ring_model2013v2}.}
\label{fig:ring_model2019_v2}
\end{figure*}

\begin{figure*}[htp]
\textbf{HD\,163296 - Ring Model (2019 June, CPs)}\par\medskip
  \centering
  \includegraphics[width=\linewidth]{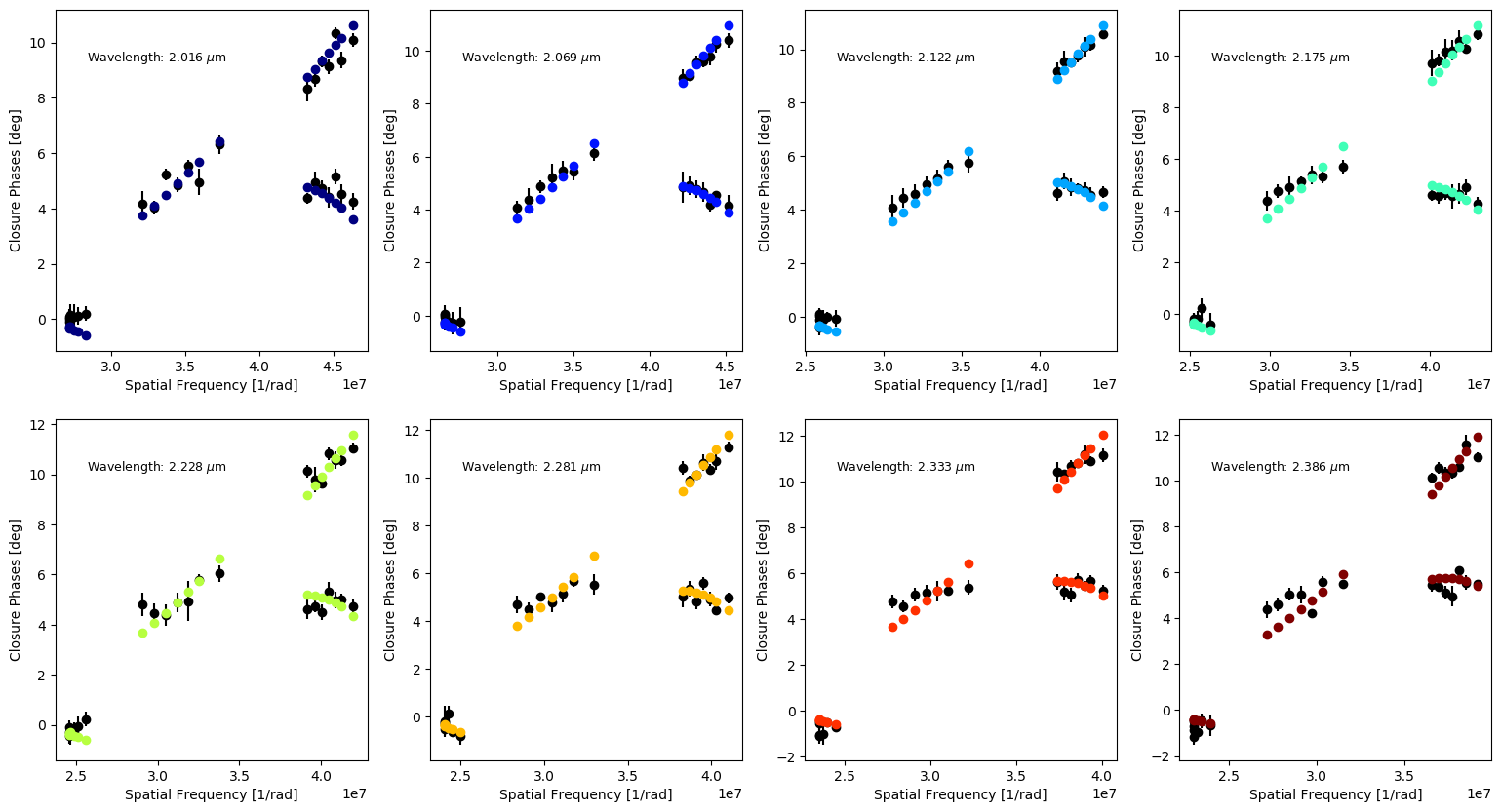}
\caption{Observations vs Ring model for the 2019 run. Panels are plotted as in Fig. \ref{fig:ring_model2013cp}.}
\label{fig:ring_model2019_cp}
\end{figure*}

\clearpage
\newpage

\section{Off-centered Gaussian model \label{sec:gaussian_model}}

This model corresponds to a point-like object (denoted with the subscript "s"), an elongated Gaussian disk (denoted with the subscript "c") and, an over-resolved component (denoted with the subscript "h"), defined by the following equation:

\begin{equation}
V(u,v) = \frac{F_{\mathrm{s}}+F_{\mathrm{c}} \times G_{\mathrm{c}}(u_{r},v_{r}) \times e^{-2\pi j (\Delta x u + \Delta y v)}   }{F_{\mathrm{s}} + F_{\mathrm{c}} + F_{\mathrm{h}}}\,.
\end{equation}

The point-like component is centered at the phase reference and the Gaussian could be displaced by a given amount $\Delta$x and $\Delta$y in Right Ascension (R.~A.) and Declination (Dec.), respectively. $F_{\mathrm{s}}$, $F_{\mathrm{c}}$ and $F_{\mathrm{h}}$ are the flux contributions between the different component, u-v are the spatial frequencies sampled with the interferometer, and  $G_{\mathrm{c}}(u_{r},v_{r})$ is the Fourier transform of a Gaussian which equals to: 

\begin{equation}
G_{\mathrm{c}}(u_{r},v_{r}) = \mathrm{exp}\left( -\frac{(\pi \Theta_{\mathrm{FWHM}} \sqrt{u_r^2 + v_r^2})^{2}}{4\mathrm{ln}2} \right) \,,    
\end{equation}

here, $\Theta_{\mathrm{FWHM}}$ is the full-width-at-half-maximum of the Gaussian.The $u_{r}$ and $v_{r}$ components are defined as in Equation \ref{eq:uv}. Tables from \ref{tab:param_model_2013} to \ref{tab:param_model_2019} display the best-fit parameters obtained for this model.

\section{Imaging \label{sec:app_imaging}}

\begin{figure*}
\textbf{2013 June/July - PIONIER Reconstructed Images (Gaussian model)}\par\medskip
  \centering
  \includegraphics[width=\linewidth]{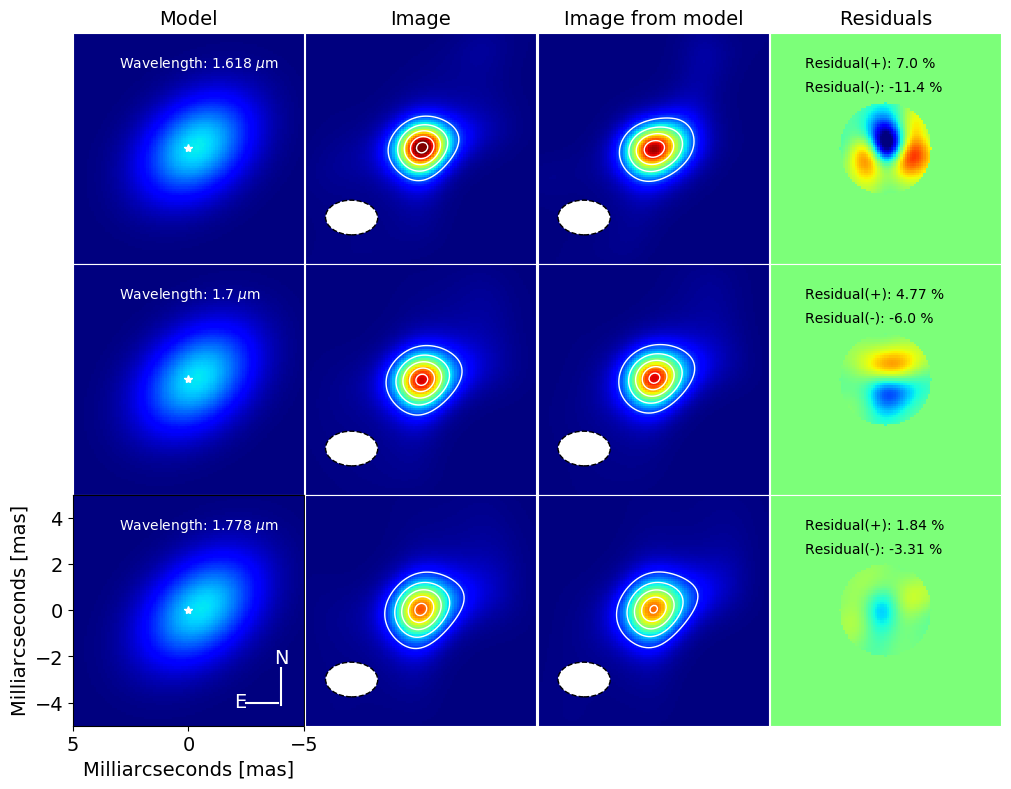}

\caption{BSMEM reconstructed images for our 2013 PIONIER epoch and the best-fit Gaussian model. The first column displays maps of the best-fit parametric models, the second column corresponds to the images recovered from the data at the corresponding epoch and wavelength. The third column shows reconstructed images from synthetic data using the indicated best-fit model. The fourth column displays the residuals between the reconstructed images from the data and the ones from the best-fit models. Wavelength for each row is labeled on the panels of the first column. The white ellipse in the second and third column correspond to the synthesized beam (at FWHM). The white contours in the images of the second and third columns correspond to 20\%, 40\%,60\%, 80\% and 95\% of the image's peak. The maximum and minimum values labeled in the fourth column show the relative percentage of the (positive and negative) residuals and the peak relative to the reconstructed image from the data.}
\label{fig:BSMEM_2013_gauss}
\end{figure*}

\begin{figure*}
\textbf{2013 June/July - PIONIER Reconstructed Images (Ring model)}\par\medskip
  \centering
  \includegraphics[width=\linewidth]{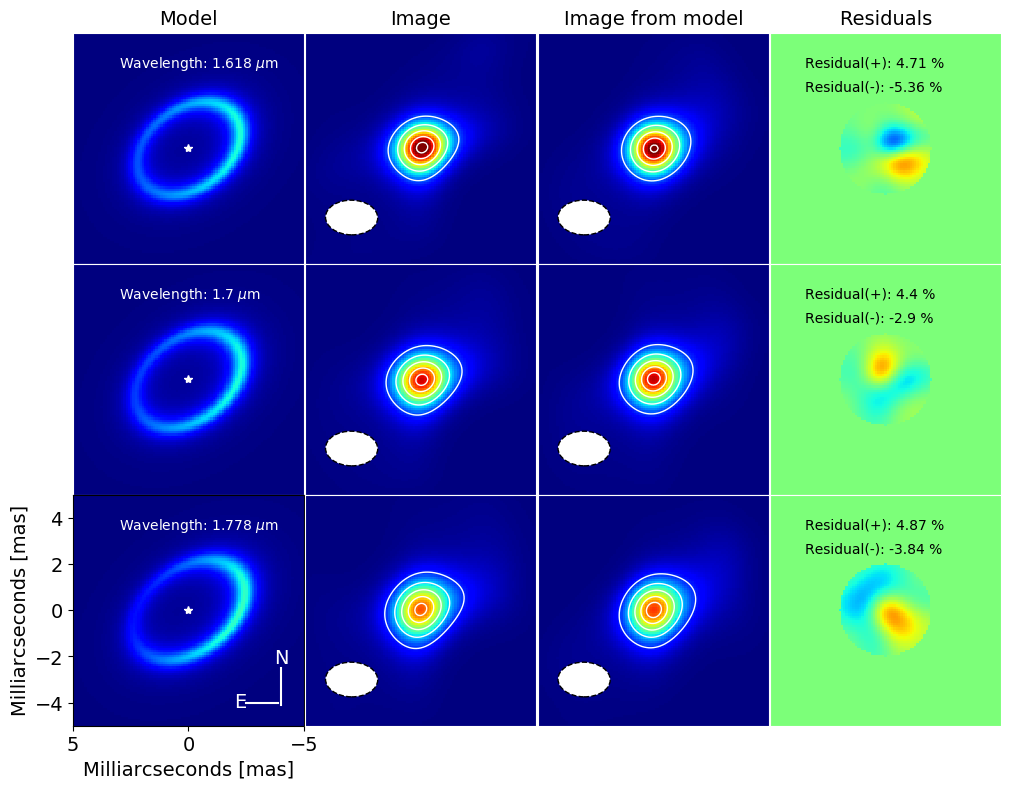}
\caption{BSMEM reconstructed images for our 2013 PIONIER epoch and the best-fit Ring model.Panels are plotted as in Fig. \ref{fig:BSMEM_2013_gauss}}
\label{fig:BSMEM_2013_ring}
\end{figure*}

\begin{figure*}
\textbf{2017 April - PIONIER Reconstructed Images (Gaussian model)}\par\medskip
  \centering
  \includegraphics[width= 14 cm]{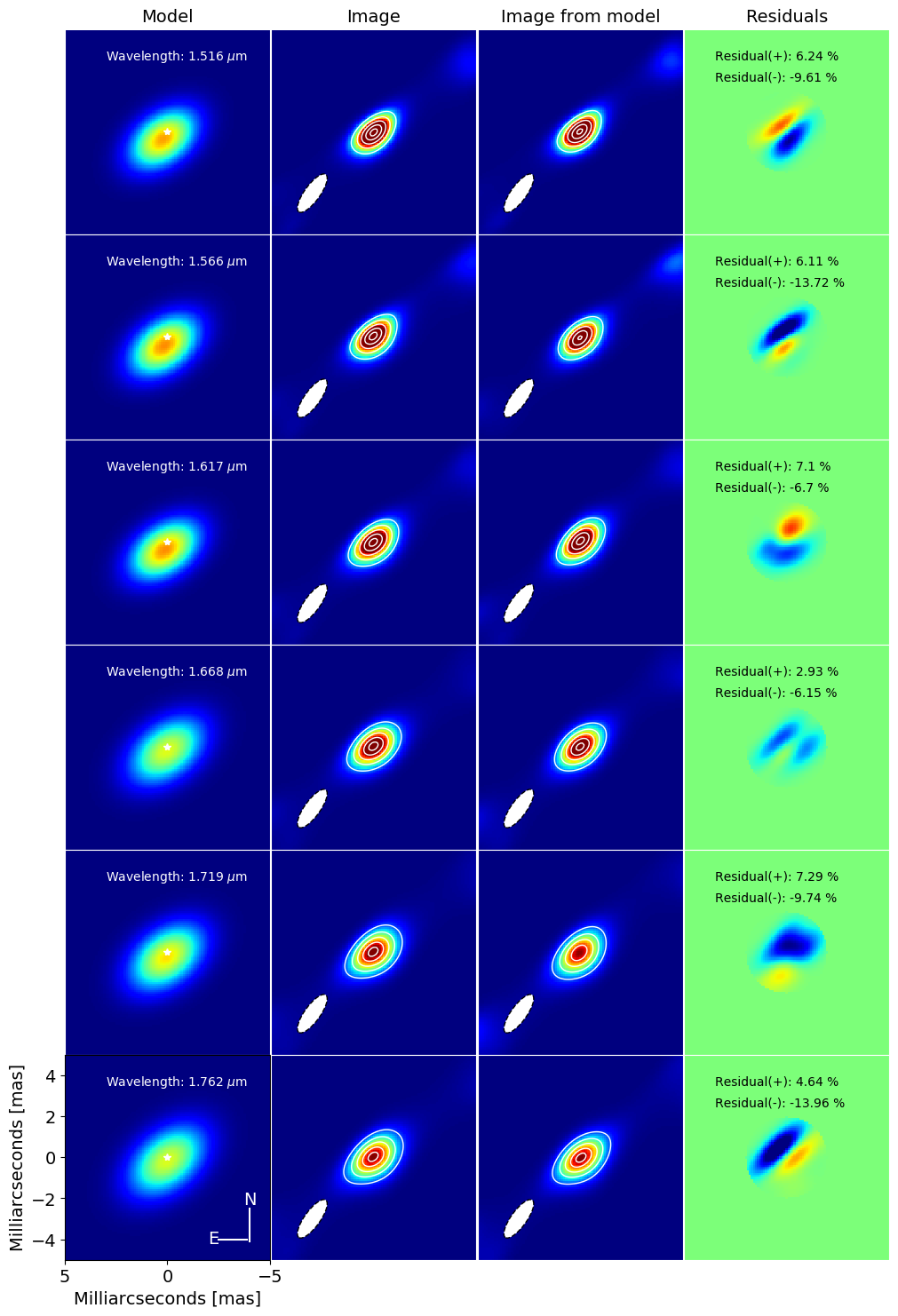}
\caption{BSMEM reconstructed images for our 2017 (April) PIONIER epoch and the best-fit Ring model.Panels are plotted as in Fig. \ref{fig:BSMEM_2013_gauss}}
\label{fig:BSMEM_2017_1_gauss}
\end{figure*}

\begin{figure*}
\textbf{2017 April - PIONIER Reconstructed Images (Ring model)}\par\medskip
  \centering
  \includegraphics[width= 14 cm]{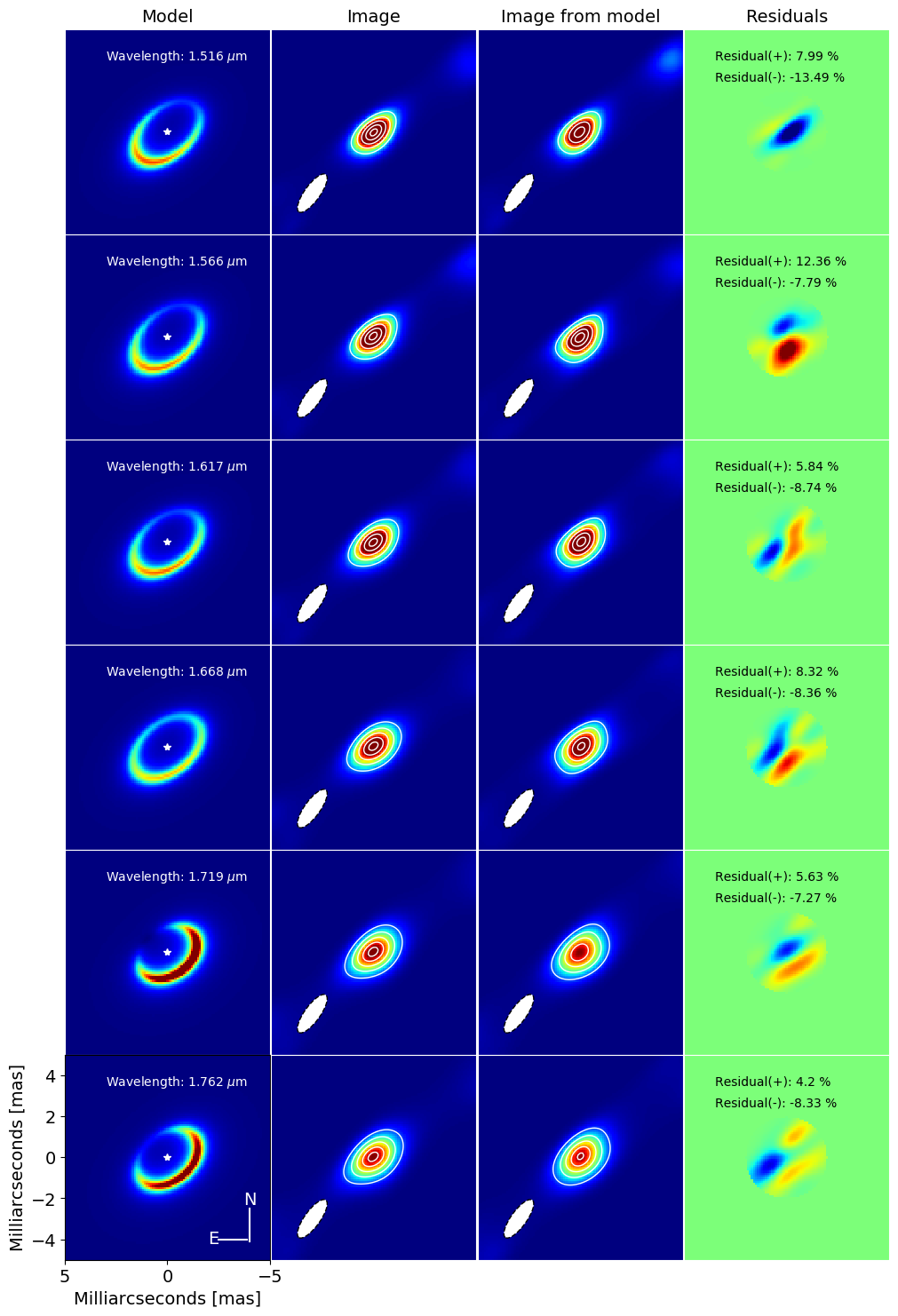}
\caption{BSMEM reconstructed images for our 2017 (April) PIONIER epoch and the best-fit Ring model.Panels are plotted as in Fig. \ref{fig:BSMEM_2013_gauss}}
\label{fig:BSMEM_2017_1_ring}
\end{figure*}

\begin{figure*}
\textbf{2018 June - GRAVITY Reconstructed Images (Gaussian model)}\par\medskip
  \centering
  \includegraphics[width= 13 cm]{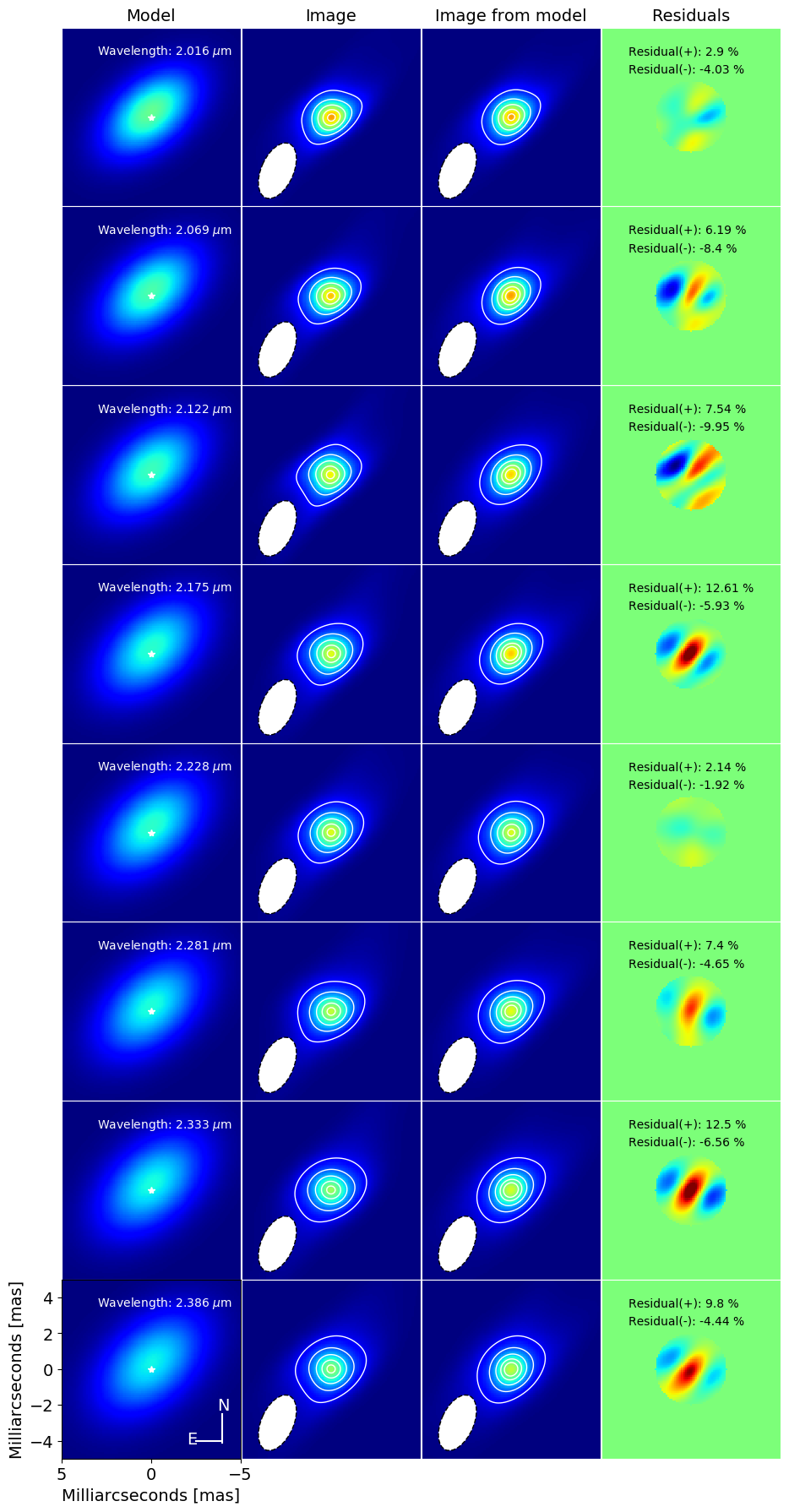}
\caption{BSMEM reconstructed images for our 2018 GRAVITY epoch and the best-fit Gaussian model.Panels are plotted as in Fig. \ref{fig:BSMEM_2013_gauss}}
\label{fig:BSMEM_2018_gauss}
\end{figure*}

\begin{figure*}
\textbf{2018 June - GRAVITY Reconstructed Images (Ring model)}\par\medskip
  \centering
  \includegraphics[width= 13 cm]{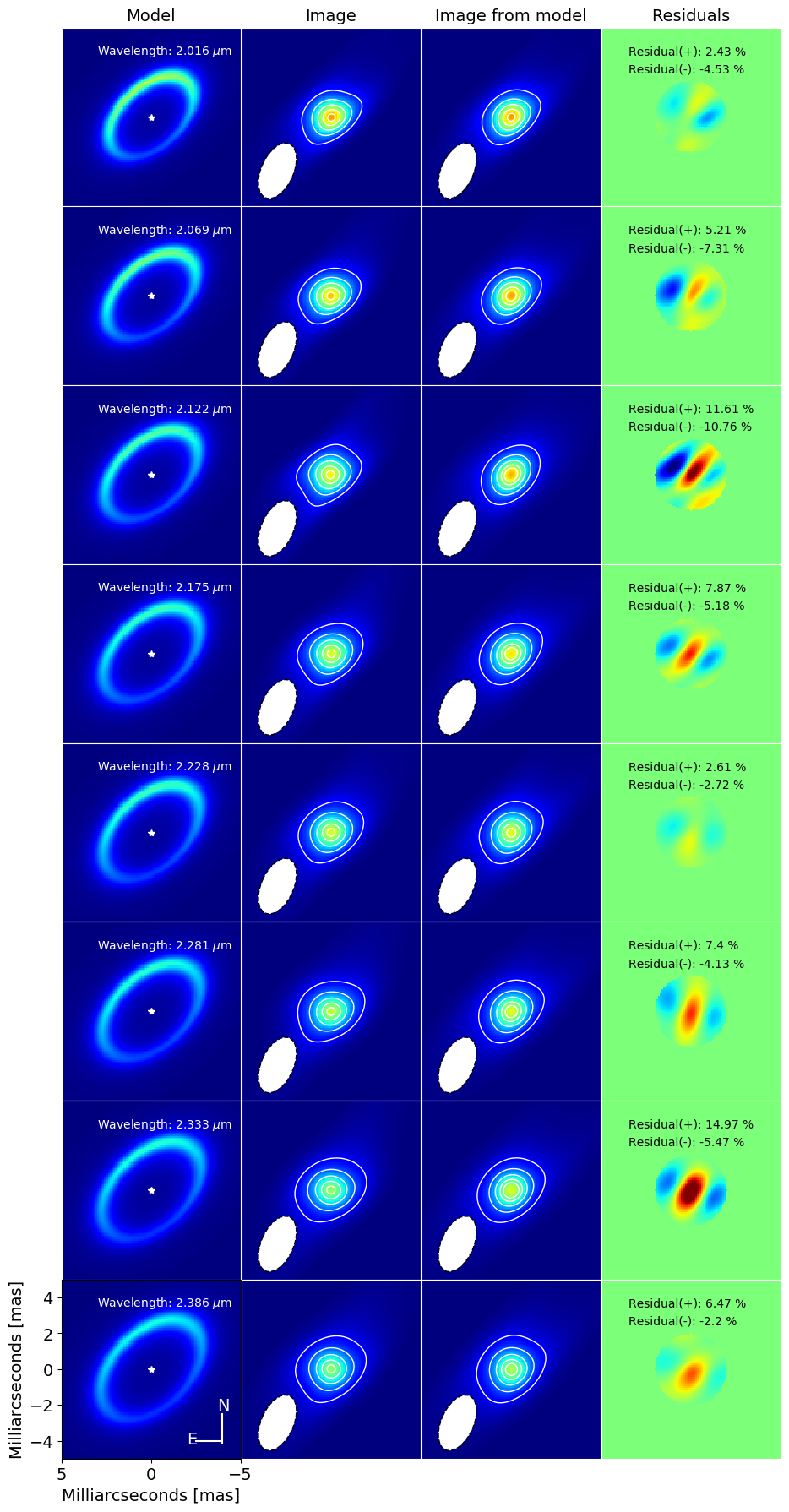}
\caption{BSMEM reconstructed images for our 2018 GRAVITY epoch and the best-fit Ring model.Panels are plotted as in Fig. \ref{fig:BSMEM_2013_gauss}}
\label{fig:BSMEM_2018_ring}
\end{figure*}

\begin{figure*}
\textbf{2019 June - GRAVITY Reconstructed Images (Gaussian model)}\par\medskip
  \centering
  \includegraphics[width= 13 cm]{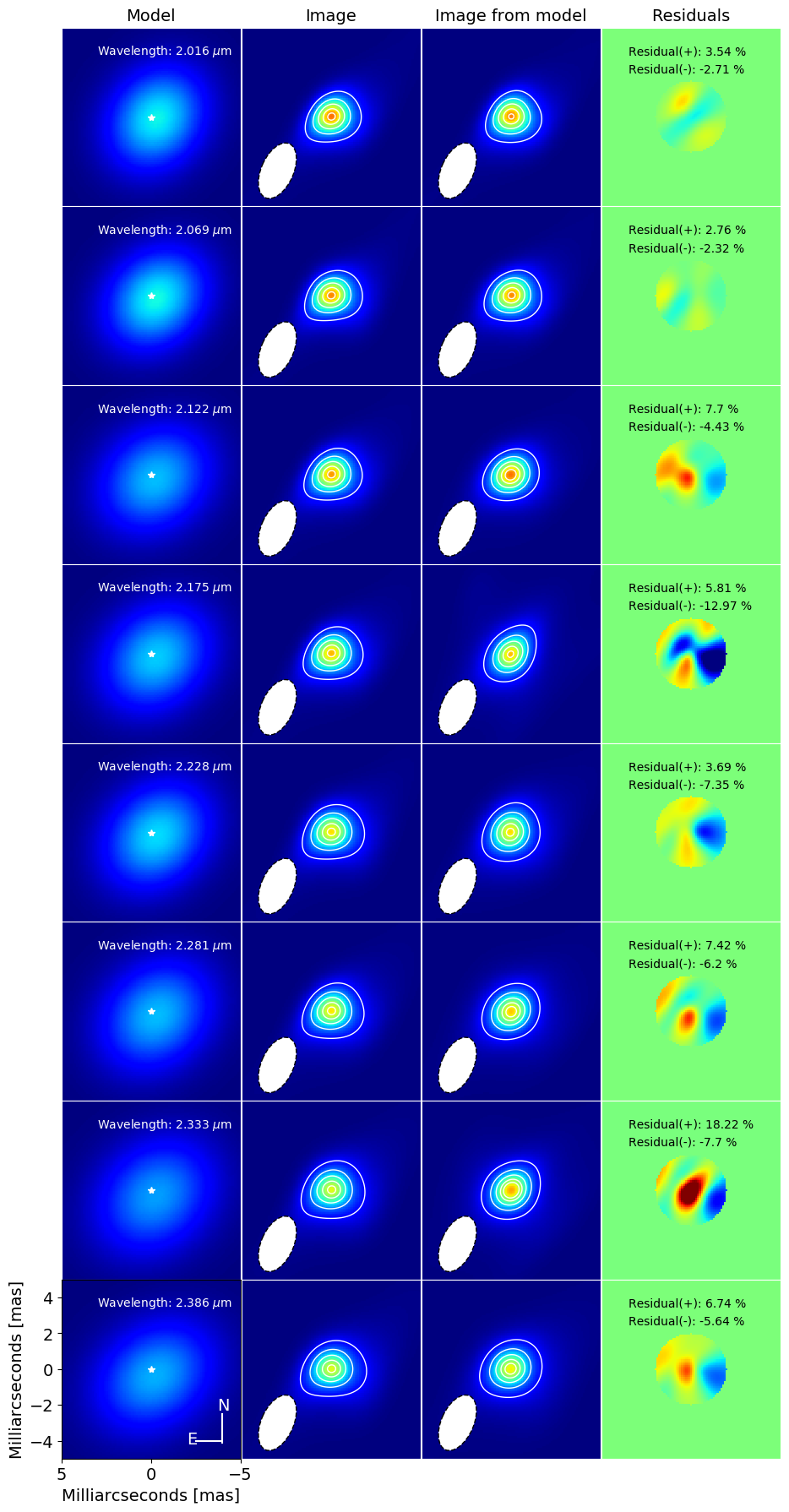}
\caption{BSMEM reconstructed images for our 2019 GRAVITY epoch and the best-fit Gaussian model.Panels are plotted as in Fig. \ref{fig:BSMEM_2013_gauss}}
\label{fig:BSMEM_2019_gauss}
\end{figure*}

\begin{figure*}
\textbf{2019 June - GRAVITY Reconstructed Images (Ring model)}\par\medskip
  \centering
  \includegraphics[width= 13 cm]{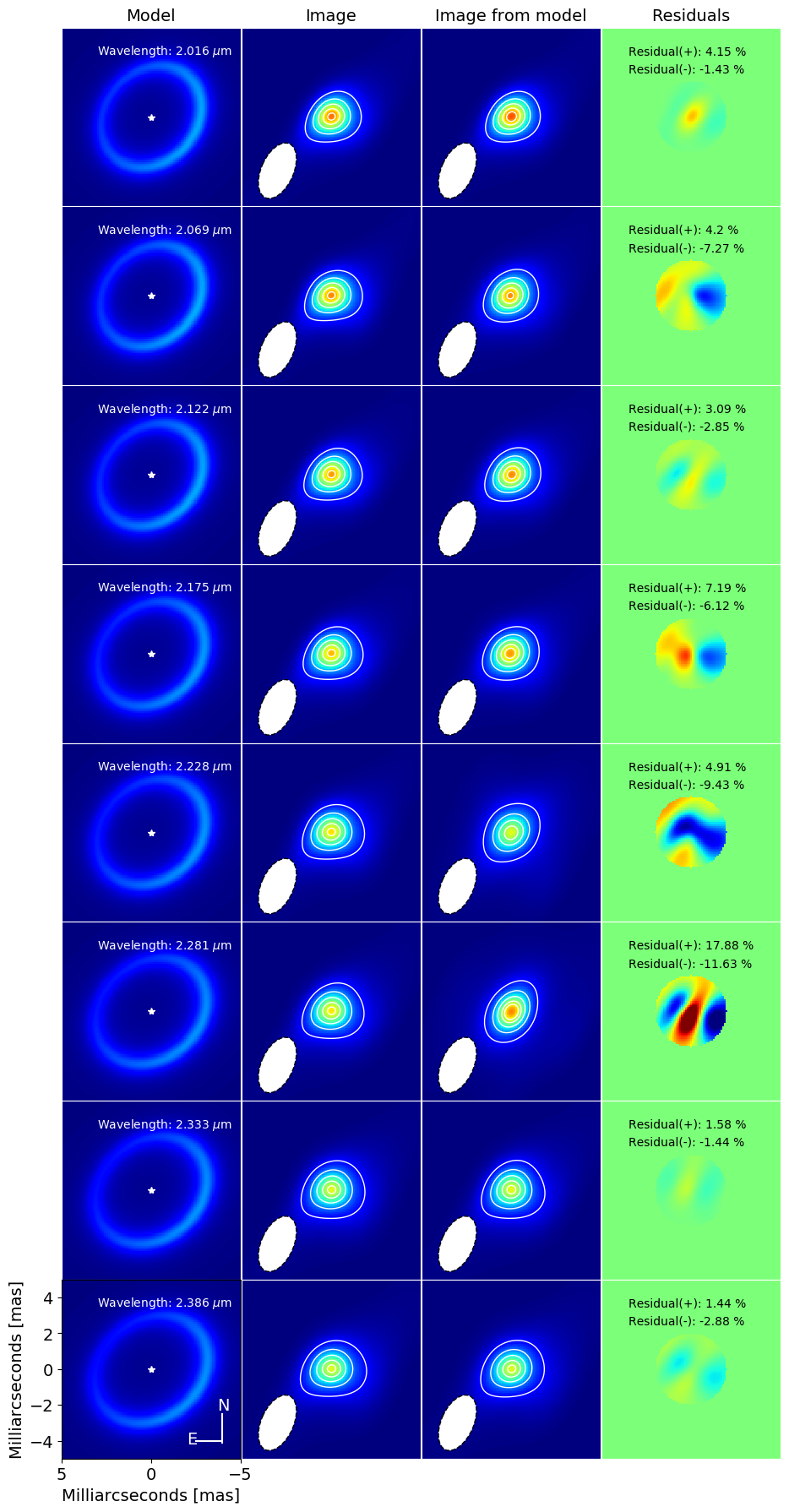}
\caption{BSMEM reconstructed images for our 2019 GRAVITY epoch and the best-fit Ring model.Panels are plotted as in Fig. \ref{fig:BSMEM_2013_gauss}}
\label{fig:BSMEM_2019_ring}
\end{figure*}

We used  BSMEM \citep{Buscher_1994, Lawson_2004_spie} to image our target. Each wavelength was imaged independently. BSMEM employs a maximum-entropy algorithm to recover the images. The code works simultaneously with CPs, V$^2$ and triple amplitudes for the reconstruction. In this case, only CPs and V$^2$ were extracted from our data, while triple amplitudes were extrapolated directly by BSMEM from the previous two observables. The code uses a Gradient Descent method to perform a regularized minimization of the general form: 

\begin{equation}
    \textbf{x} = \arg\min_{x}\,\,\Bigl\{\,\chi^2 + \alpha R(x)\Bigr\}\,,
\end{equation}

\noindent where \textbf{x} is the sought most-probable image that reproduces our data; $\chi^2$ is the negative log-likelihood between the synthetic observables retrieved from a model image and our data; $R(x)$ is the prior term which includes the information that is known about the source and; $\alpha$ is an hyperparameter that weights between the likelihood and the prior. BSMEM uses the Gull-Skilling entropy as prior:

\begin{equation}
R(\textbf{x}) = \sum_n [x_n\mathrm{log}(x_n/\overline{x}_n)-x_n+\overline{x}_n]\,,
\end{equation}

where $x_n$ is the value of the image at pixel $n$ and $\overline{x}$ is the initial image, which is recovered in absence of data. This regularizer enforces positivity in the image and favors smooth extended structures over sparse-compact ones;  BSMEM selects automatically the hyper-parameter value. Images recovered here uses a pixel scale of 0.1 mas and they are recovered over a pixel grid of 501 $\times$ 501 pixels. To not over-regularize the reconstruction with the initial conditions, the starting image was set to be a Gaussian (FWHM = 4 mas) centered in the middle of the pixel grid. Images are presented as recovered from BSMEM and no subsequent smoothing convolution was done. All the reconstructions converged, however, we noticed a significant increased in the $\chi^2$ for the 2018 epoch, compared with the other four epochs. Still, the observables of the 2018 epoch are well reproduced by the images. Additionally to the recovered images obtained from the data, we also reconstructed images using observables generated from the best-fit Ring and Gaussian models, while keeping the SNR statistics of the data. These images serve us to compare the morphology observed in the reconstructed images and the one obtained from the parametric models, after being recovered using the same setup and imaging algorithm. The images from the data per wavelength and the images from the parametric models can be consulted from Figure \ref{fig:BSMEM_2013_gauss} to \ref{fig:BSMEM_2019_ring}. The $\chi^{2}$ of the images, the hyper-parameter value and the number of iterations of the reconstructions are listed from Table \ref{tab:param_model_2013} to \ref{tab:param_model_2019}

The reconstructed images from the observables extracted from the best-fit parametric models are quite similar to the reconstructed images from the data. Our imaging experiment does not show a clear preference over the two parametric models to discriminate which one could better reproduce the image from the data. While, on average, residuals remain at 10\% of the peak in the image, we noticed that for some spectral channels, we observed residuals as large as 20\% when we compare the reconstructed images and the best-fit models. We suspect that this difference is caused by asymmetric structures in the morphology of the target not traced by our parametric models. To have more quantitative estimates on these structures, a compelling u-v coverage is required to do a high-fidelity image of the source.

\begin{table*}[htp]
\caption[]{PIONIER (2013 June/July) - Best-fit parameters of the parametric models}
\label{tab:param_model_2013}
\centering
\rotatebox{0}{
\begin{tabular}{l c c c } 
\hline \hline
\multicolumn{4}{c}{\textbf{GAUSSIAN MODEL}} \\
\hline
Wavelengths [microns] & 1.618 & 1.7 & 1.778 \\
\hline
$\Theta_{\mathrm{FWHM}}$ [mas] &            3.03 $\pm$ 0.1 &  3.07  $\pm$ 0.08 & 3.05 $\pm$ 0.09 \\
$F_{\mathrm{s}}$ &                   0.31 $\pm$ 0.01 &  0.29 $\pm$ 0.01  & 0.26 $\pm$ 0.01 \\
$F_{\mathrm{c}}$ &                    0.63 $\pm$ 0.01 &  0.65 $\pm$ 0.01  & 0.67 $\pm$ 0.01 \\
$F_{\mathrm{h}}$       & 0.06 $\pm$ 0.01  &   0.06 $\pm$ 0.01  & 0.07 $\pm$ 0.01  \\
PA [deg] &                  140.6 $\pm$ 3.5 &  139.8 $\pm$ 1.9  & 139.0 $\pm$ 1.7 \\
i [deg]  &                  44.3 $\pm$ 1.9  &  45.6 $\pm$ 1.4   & 47.6 $\pm$ 1.5   \\
$\Delta$x [mas] &           -0.35 $\pm$ 0.09  & -0.39 $\pm$ 0.06  & -0.53 $\pm$ 0.07 \\
$\Delta$y [mas] &           0.07 $\pm$ 0.07  & 0.05 $\pm$ 0.05  & 0.06 $\pm$ 0.05  \\
$\chi$2 & 2.4 & 3.0 & 4.4  \\
\hline \hline
\multicolumn{4}{c}{\textbf{RECONSTRUCTED IMAGES}} \\
\hline
Wavelengths [microns] & 1.618 & 1.7 & 1.778 \\
\hline
Hyper-parameter ($\alpha$)   & 231.7  &  364     & 407  \\
Iterations                   & 37     &  36      & 34 \\
$\chi^2$                     & 1.01   &  1.01    & 1.0  \\
\hline
\end{tabular}
}
\end{table*}

\begin{table*}[htp]
\caption[]{PIONIER (2017 April) - Best-fit parameters of the parametric models}
\label{tab:param_model_2017_1}
\centering
\rotatebox{0}{
\begin{tabular}{l c c c c c c } 
\hline \hline
\multicolumn{7}{c}{\textbf{GAUSSIAN MODEL}} \\
\hline
Wavelengths [microns] & 1.518 & 1.567 & 1.617 & 1.67 & 1.72 & 1.763 \\
\hline
$\Theta_{\mathrm{FWHM}}$ [mas] &            1.8 $\pm$ 0.1    &  1.85  $\pm$ 0.1   & 1.84 $\pm$ 0.13  & 2.1 $\pm$ 0.12     & 2.04 $\pm$ 0.15 & 2.05 $\pm$ 0.15 \\
$F_{\mathrm{s}}$ &                    0.29 $\pm$ 0.02   &  0.26 $\pm$ 0.03   & 0.23 $\pm$ 0.03  & 0.24 $\pm$ 0.03    & 0.20 $\pm$ 0.03 & 0.20 $\pm$ 0.03 \\
$F_{\mathrm{c}}$ &                    0.53 $\pm$ 0.01   &  0.56 $\pm$ 0.02   & 0.56 $\pm$ 0.02  & 0.56 $\pm$ 0.02    & 0.60 $\pm$ 0.03 & 0.60 $\pm$ 0.02 \\
$F_{\mathrm{h}}$       & 0.18 $\pm$ 0.01  &  0.18 $\pm$ 0.01   & 0.20 $\pm$ 0.01  & 0.20 $\pm$ 0.02    & 0.20 $\pm$ 0.009  & 0.20 $\pm$ 0.02 \\
PA [deg] &                  142.7 $\pm$ 1.4  &  143.0 $\pm$ 1.2   & 144.9 $\pm$ 1.6  & 142.6 $\pm$ 1.7    & 142.5$\pm$ 2.2 & 142.5 $\pm$ 2.2  \\
i [deg]  &                  49.9 $\pm$ 0.8   &  50.6 $\pm$ 0.7    & 51.1 $\pm$ 0.9   & 50.5 $\pm$ 0.8     & 49.8 $\pm$ 1.1  & 49.8 $\pm$ 1.1   \\
$\Delta$x [mas] &           0.20 $\pm$ 0.03  & 0.17 $\pm$ 0.04    & 0.12 $\pm$ 0.04  & 0.02 $\pm$ 0.04    & 0.03 $\pm$ 0.05 & 0.03 $\pm$ 0.05  \\
$\Delta$y [mas] &           -0.35 $\pm$ 0.06 & -0.42 $\pm$ 0.06   & -0.38 $\pm$ 0.08  & -0.17 $\pm$ 0.06  & -0.23 $\pm$ 0.08 & -0.23 $\pm$ 0.08  \\
$\chi$2 & 1.1 & 1.6 & 2.6 & 3.3 & 4.4 & 4.4  \\
\hline \hline
\multicolumn{7}{c}{\textbf{RECONSTRUCTED IMAGES}} \\
\hline
Wavelengths [microns] & 1.518 & 1.567 & 1.617 & 1.67 & 1.72 & 1.763 \\
\hline
Hyper-parameter ($\alpha$) & 146   &  231   &  203   & 199  & 204   & 188 \\
Iterations                 & 93    &  122   &  89    & 82   &  88   & 86 \\
$\chi^2$                   & 0.97  &  1.01  &  1.02  & 1.01 & 0.99  & 1.00 \\
\hline
\end{tabular}
}
\end{table*}

\begin{table*}[htp]
\caption[]{PIONIER (2017 August) - Best-fit parameters of the parametric models}
\label{tab:param_model_2017_2}
\centering
\rotatebox{0}{
\begin{tabular}{l c c c c c c } 
\hline \hline
\multicolumn{7}{c}{\textbf{GAUSSIAN MODEL}} \\
\hline
Wavelengths [microns] & 1.518 & 1.567 & 1.617 & 1.67 & 1.72 & 1.763 \\
\hline
$\Theta_{\mathrm{FWHM}}$ [mas] &     1.23 $\pm$ 0.07  &  1.3  $\pm$ 0.06  & 1.3 $\pm$ 0.05    & 1.3 $\pm$ 0.05   & 1.36 $\pm$ 0.05 & 1.33 $\pm$ 0.06 \\
$F_{\mathrm{s}}$ &                   0.27 $\pm$ 0.01    &  0.25 $\pm$ 0.01  & 0.21 $\pm$ 0.01   & 0.19 $\pm$ 0.01   & 0.18 $\pm$ 0.009 & 0.19 $\pm$ 0.007 \\
$F_{\mathrm{c}}$ &                    0.49 $\pm$ 0.02   &  0.52 $\pm$ 0.01  & 0.55 $\pm$ 0.01   & 0.55 $\pm$ 0.02   & 0.56 $\pm$ 0.01 & 0.64 $\pm$ 0.006 \\
$F_{\mathrm{h}}$       & 0.24 $\pm$ 0.02  &  0.23 $\pm$ 0.01  & 0.24 $\pm$ 0.01   & 0.26 $\pm$ 0.02   & 0.26 $\pm$ 0.01  & 0.17 $\pm$ 0.008 \\
PA [deg] &                  141.0 $\pm$ 1.9  &  142.1 $\pm$ 1.4  & 140.8 $\pm$ 1.3   & 140.1 $\pm$ 1.4   & 141.9$\pm$ 1.4 & 138.3 $\pm$ 0.6  \\
i [deg]  &                  59.8 $\pm$ 1.3   &  59.1 $\pm$ 1.0   & 58.5 $\pm$ 0.8    & 58.4 $\pm$ 0.9    & 58.5 $\pm$ 0.9  & 52.2 $\pm$ 0.3   \\
$\Delta$x [mas] &           0.05 $\pm$ 0.03  & 0.06 $\pm$ 0.02   & 0.12 $\pm$ 0.02   & 0.13 $\pm$ 0.03   & 0.19 $\pm$ 0.02 & 0.01 $\pm$ 0.02  \\
$\Delta$y [mas] &           -0.24 $\pm$ 0.1  & -0.27 $\pm$ 0.08  & -0.25 $\pm$ 0.08  & -0.23 $\pm$ 0.1   & -0.4 $\pm$ 0.08 & 0.29 $\pm$ 0.03  \\
$\chi$2 & 2.5 & 2.5 & 3.7 & 5.3 & 3.0 & 2.7  \\
\hline
\hline
\multicolumn{7}{c}{\textbf{RECONSTRUCTED IMAGES}} \\
\hline
Wavelengths [microns] & 1.518 & 1.567 & 1.617 & 1.67 & 1.72 & 1.763 \\
\hline
Hyper-parameter ($\alpha$) & 166   &  222   &  198   & 194  & 213   & 180 \\
Iterations                 & 97    &  122   &  127    & 131   &  111   & 127 \\
$\chi^2$                   & 1.01  &  1.01  &  1.0  & 1.02 & 1.02  & 0.99 \\
\hline
\end{tabular}
}
\end{table*}

\begin{table*}[htp]
\caption[]{GRAVITY (2018 June) - Best-fit parameters of the parametric models}
\label{tab:param_model_2018}
\centering
\rotatebox{90}{
\begin{tabular}{l c c c c c c c c } 
\hline \hline
\multicolumn{9}{c}{\textbf{GAUSSIAN MODEL}} \\
\hline
Wavelengths [microns] & 2.016 & 2.069 & 2.122 & 2.175 & 2.228 & 2.281 & 2.333 & 2.386 \\
\hline
$\Theta_{\mathrm{FWHM}}$ [mas] &            2.87 $\pm$ 0.06 &  3.03  $\pm$ 0.06 & 3.14 $\pm$ 0.05  & 3.26 $\pm$ 0.06   & 3.27 $\pm$ 0.06 & 3.33 $\pm$ 0.06   & 3.40 $\pm$ 0.06  & 3.49 $\pm$ 0.08 \\
$F_{\mathrm{s}}$ &                   0.20 $\pm$ 0.007 &  0.21 $\pm$ 0.007  & 0.20 $\pm$ 0.006  & 0.21 $\pm$ 0.006   & 0.20 $\pm$ 0.007 & 0.19 $\pm$ 0.007 & 0.19 $\pm$ 0.007  & 0.18 $\pm$ 0.009 \\
$F_{\mathrm{c}}$ &                    0.60 $\pm$ 0.008 &  0.61 $\pm$ 0.007  & 0.63 $\pm$ 0.006  & 0.63 $\pm$ 0.006   & 0.64 $\pm$ 0.006 & 0.64 $\pm$ 0.006 & 0.64 $\pm$ 0.006  & 0.65 $\pm$ 0.007 \\
$F_{\mathrm{h}}$       & 0.2 $\pm$ 0.01  &   0.18 $\pm$ 0.01  & 0.17 $\pm$ 0.009  & 0.16 $\pm$ 0.008   & 0.16 $\pm$ 0.009  & 0.17 $\pm$ 0.008 & 0.17 $\pm$ 0.008  & 0.16 $\pm$ 0.009 \\
PA [deg] &                  139.9 $\pm$ 0.7 &  139.6 $\pm$ 0.6  & 139.2 $\pm$ 0.6  & 138.6 $\pm$ 0.6   & 137.9$\pm$ 0.5 & 138.3 $\pm$ 0.6 & 138.0 $\pm$ 0.5  & 137.3 $\pm$ 0.6 \\
i [deg]  &                  53.3 $\pm$ 0.4  &  53.0 $\pm$ 0.4   & 52.7 $\pm$ 0.3   & 52.3 $\pm$ 0.3    & 52.4 $\pm$ 0.3  & 52.2 $\pm$ 0.3  & 51.6 $\pm$ 0.3   & 51.0 $\pm$ 0.3 \\
$\Delta$x [mas] &           0.03 $\pm$ 0.02  & 9e-3 $\pm$ 0.02  & -3e-3 $\pm$ 0.02 & -0.03 $\pm$ 0.01 & -0.03 $\pm$ 0.02 & 0.01 $\pm$ 0.02 & -0.04 $\pm$ 0.02 & -0.04 $\pm$ 0.03  \\
$\Delta$y [mas] &           0.25 $\pm$ 0.03  & 0.28 $\pm$ 0.03  & 0.30 $\pm$ 0.03  & 0.33 $\pm$ 0.02   & 0.35 $\pm$ 0.03 & 0.29 $\pm$ 0.03 & 0.36 $\pm$ 0.04  & 0.37 $\pm$ 0.04 \\
$\chi$2 & 2.9 & 8.0 & 11.4 & 5.6 & 4.4 & 5.0 & 5.3 & 3.6 \\
\hline
\hline
\multicolumn{9}{c}{\textbf{RECONSTRUCTED IMAGES}} \\
\hline
Wavelengths [microns] & 2.016 & 2.069 & 2.122 & 2.175 & 2.228 & 2.281 & 2.333 & 2.386 \\
\hline
Hyper-parameter ($\alpha$) & 200   &  201   &  200   & 200  & 201   & 200 & 201 & 199 \\
Iterations                 & 33    &  42   &  46    & 42   &  39   & 40 &  43 & 37 \\
$\chi^2$                   & 1.65  &  3.07  &  3.56  & 4.0 & 4.8  & 5.6 & 5.9 & 4.5 \\
\hline
\end{tabular}
}
\end{table*}

\begin{table*}[htp]
\caption[]{GRAVITY (2019 June) - Best-fit parameters of the parametric models}
\label{tab:param_model_2019}
\centering
\rotatebox{90}{
\begin{tabular}{l c c c c c c c c } 
\hline \hline
\multicolumn{9}{c}{\textbf{GAUSSIAN MODEL}} \\
\hline
Wavelengths [microns] & 2.016 & 2.069 & 2.122 & 2.175 & 2.228 & 2.281 & 2.333 & 2.386 \\
\hline
$\Theta_{\mathrm{FWHM}}$ [mas]            & 4.03 $\pm$ 0.11   & 3.99 $\pm$ 0.08  & 4.37 $\pm$ 0.08   & 4.26 $\pm$ 0.07   & 4.20 $\pm$ 0.06  & 4.40 $\pm$ 0.06  & 4.63 $\pm$ 0.06  & 4.4 $\pm$ 0.09 \\
$F_{\mathrm{s}}$                     & 0.26 $\pm$ 0.01   & 0.25 $\pm$ 0.007  & 0.28 $\pm$ 0.004   & 0.26 $\pm$ 0.006   & 0.24 $\pm$ 0.006  & 0.25 $\pm$ 0.006  & 0.27 $\pm$ 0.004  & 0.25 $\pm$ 0.007 \\
$F_{\mathrm{c}}$                     & 0.66 $\pm$ 0.008    & 0.66 $\pm$ 0.007  & 0.66 $\pm$ 0.007   & 0.66 $\pm$ 0.006   & 0.68 $\pm$ 0.006  & 0.67 $\pm$ 0.005  & 0.68 $\pm$ 0.003  & 0.68 $\pm$ 0.006 \\
$F_{\mathrm{h}}$      & 0.08 $\pm$ 0.01    & 0.09 $\pm$ 0.009  & 0.06 $\pm$ 0.009   & 0.08 $\pm$ 0.007   & 0.08 $\pm$ 0.007  & 0.08 $\pm$ 0.005  & 0.06 $\pm$ 0.005  & 0.07 $\pm$ 0.007 \\
PA [deg]                  & 127.5 $\pm$ 2.3   & 135.7 $\pm$ 1.6  & 141.5 $\pm$ 2.0   & 138.1 $\pm$ 1.6   & 135.9 $\pm$ 1.0  & 137.0 $\pm$ 1.2  & 139.0 $\pm$ 0.7  & 144.0 $\pm$ 1.4 \\
i [deg]                   & 34.1 $\pm$ 1.2    &  35.3 $\pm$ 0.6  & 34.3 $\pm$ 1.1    & 35.1 $\pm$ 0.7    & 36.1 $\pm$ 0.4   & 35.5 $\pm$ 0.6   & 35.9 $\pm$ 0.4   & 38.5 $\pm$ 0.8 \\
$\Delta$x [mas]           & -0.3 $\pm$ 0.02   & -0.3 $\pm$ 0.02  & -0.23 $\pm$ 0.01  & -0.29 $\pm$ 0.01  & -0.32 $\pm$ 0.02 & -0.31 $\pm$ 0.01 & -0.26 $\pm$ 0.01 & -0.23 $\pm$ 0.01  \\
$\Delta$y [mas]           & -0.1 $\pm$ 0.03   & - 0.2 $\pm$ 0.02 & -0.25 $\pm$ 0.02  & -0.21 $\pm$ 0.01  & -0.21 $\pm$ 0.02 & -0.24 $\pm$ 0.02 & -0.32 $\pm$ 0.01 & -0.37 $\pm$ 0.02 \\
$\chi$2 & 2.7 & 1.5 & 1.9 & 1.1 & 1.5 & 1.3 & 1.2 & 1.7 \\
\hline \hline
\multicolumn{9}{c}{\textbf{RECONSTRUCTED IMAGES}} \\
\hline
Wavelengths [microns] & 2.016 & 2.069 & 2.122 & 2.175 & 2.228 & 2.281 & 2.333 & 2.386 \\
\hline
Hyper-parameter ($\alpha$) & 707   &  706   &  968   & 861  & 601   & 668 & 790 & 379 \\
Iterations                 & 18    &  30   &  27    & 29   &  28   & 26 &  28 & 32 \\
$\chi^2$                   & 0.98  &  0.99  &  1.0  & 1.01 & 0.99  & 1.0 & 1.01 & 1.01 \\
\hline
\end{tabular}
}
\end{table*}

\newpage

\begin{figure*}
\textbf{2017 Aug. - PIONIER Reconstructed Images (Gaussian model)}\par\medskip
  \centering
  \includegraphics[width= 14 cm]{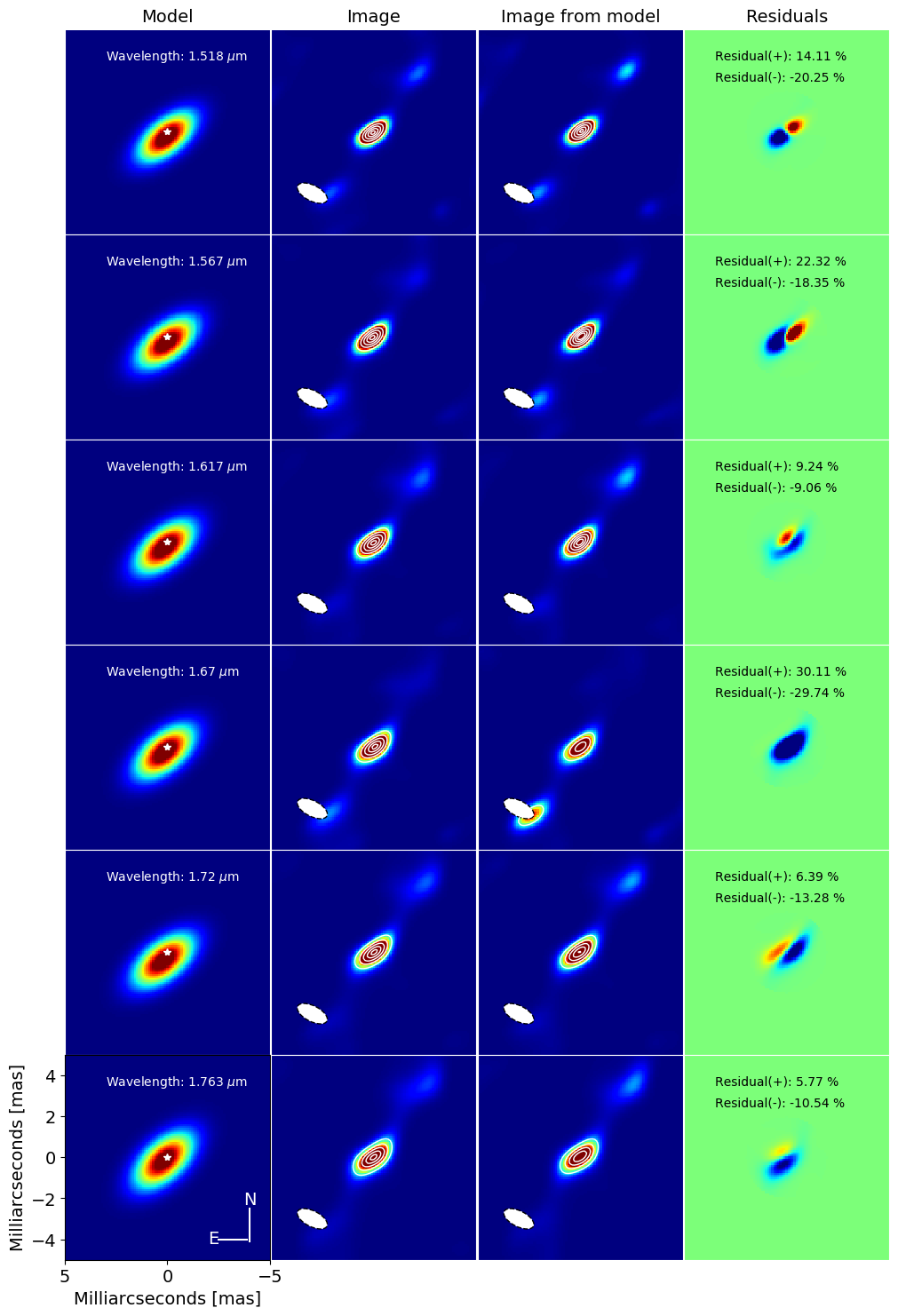}
\caption{BSMEM reconstructed images for our 2017 (Aug.) PIONIER epoch and the best-fit Gaussian model.Panels are plotted as in Fig. \ref{fig:BSMEM_2013_gauss}}
\label{fig:BSMEM_2017_2_gauss}
\end{figure*}

\begin{figure*}
\textbf{2017 Aug. - PIONIER Reconstructed Images (Ring model)}\par\medskip
  \centering
  \includegraphics[width= 14 cm]{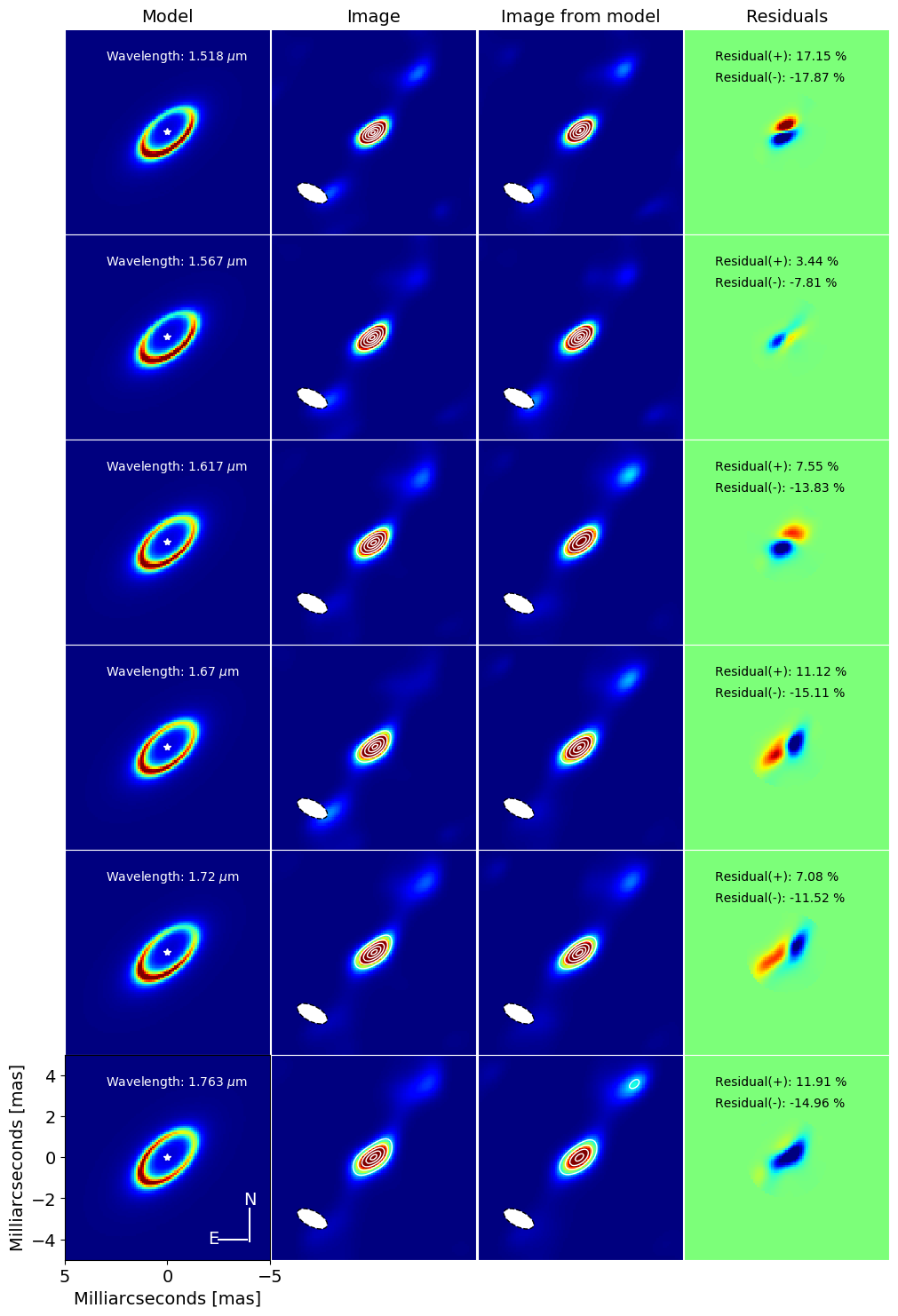}
\caption{BSMEM reconstructed images for our 2017 (Aug.) PIONIER epoch and the best-fit Ring model.Panels are plotted as in Fig. \ref{fig:BSMEM_2013_gauss}}
\label{fig:BSMEM_2017_2_gauss}
\end{figure*}

\end{appendix}
\end{document}